\documentclass[10pt,superscriptaddress,twocolumn,amsmath,amssymb,aps,prl,showpacs]{revtex4-1}
\usepackage{mathrsfs}
\usepackage{graphicx}
\usepackage{dcolumn}
\usepackage{bm}
\usepackage{amssymb}
\usepackage{amsmath}
\usepackage{paralist}
\usepackage{float}
\usepackage{mdframed}
\usepackage{booktabs}
\usepackage{mathtools} 
\usepackage{graphicx}
\usepackage[normalem]{ulem}
\usepackage[colorlinks,linkcolor=blue,anchorcolor=blue,citecolor=green]{hyperref} 
\usepackage{cleveref}
\usepackage{url}
\usepackage{color}
\usepackage{verbatim}

\newcommand{\e}{\mathrm{e}}
\newcommand{\Li}{\mathrm{Li}}
\newcommand{\mi}{\mathrm{i}}
\renewcommand{\d}{\mathrm{d}}
\renewcommand{\exp}{\mathrm{exp}}
\newcommand{\ph}{\mathrm{ph}}
\newcommand{\dr}{\mathrm{dr}}
\newcommand{\eff}{\mathrm{eff}}
\newcommand{\no}{\nonumber}

\begin{document}
	
\title{Quantum transport in 1D  Hubbard model: Drude weights and Seebeck effect  }
	
\author{Jia-Jia Luo}
\affiliation{Innovation Academy for Precision Measurement Science and Technology, Chinese Academy of Sciences, Wuhan 430071, China}
\affiliation{University of Chinese Academy of Sciences, Beijing 100049, China.}
	
\author{Sagarika Basak}
\affiliation{Department of Physics and Astronomy, and Smalley-Curl Institute,
Rice University, Houston, Texas 77251-1892, USA}
\affiliation{Homer L. Dodge Department of Physics and Astronomy, The University of Oklahoma, 440 W. Brooks Street, Norman, Oklahoma 73019, USA}

\author{Han Pu}
\email[]{hpu@rice.edu}
\affiliation{Department of Physics and Astronomy, and Smalley-Curl Institute,
Rice University, Houston, Texas 77251-1892, USA}

\author{Xi-Wen Guan}
\email[]{xiwen.guan@anu.edu.au}
\affiliation{Innovation Academy for Precision Measurement Science and Technology, Chinese Academy of Sciences, Wuhan 430071, China}
\affiliation{Hefei National Laboratory, Hefei 230088, People’s Republic of China}
\affiliation{Department of Fundamental and Theoretical Physics, Research School of Physics,
Australian National University, Canberra, ACT 0200, Australia}
	
	
\date{\today}
	
\begin{abstract}
The Drude weight (DW) is an essential quantity that characterizes the quantum transport properties of many-body systems. 
 However, a rigorous understanding and exact computation of DWs, particularly for strongly correlated systems with doping, still remain elusive.
  In this Letter, taking advantage of the quantum integrability, we calculate exactly the DWs and Seebeck effect (SE)  for generic filling factor in one-dimensional (1D) Fermi-Hubbard model with arbitrary interaction strengths and magnetic fields. We build up its intrinsic connection to the Luttinger parameters, and derive universal scaling laws for DWs across phase transitions. Our results provide a deep understanding of mutual influences in transport between the spin and the charge degrees of freedom, showing a counterintuitive subtle spin-charge coupling effect and uncovering the microscopic origin of the (spin) Seebeck effects in thermal conductivity. Finally, we propose an experimental protocol to measure the DWs in ultracold atomic systems. 

\end{abstract}
	
\maketitle

Quantum transport often reveals many intrinsic properties of a many-body system and, hence, has attracted great attention over several decades. Transport dynamics is typically initiated when a system is subjected to a potential gradient. Generically, we can write
\begin{equation}
	\left(\begin{array}{cc}J^c \\ J^s \\ J^{e}\end{array}\right)=\left (\begin{array}{cccc}
	\sigma^{cc} &\sigma^{cs}   & \sigma^{ce} \\
	\sigma^{sc} &\sigma^{ss} & \sigma^{se} \\
	\sigma^{ec} & \sigma^{es} &\sigma^{ee} \\
	\end{array}\right)\left(\begin{array}{cc}\nabla \mu \\ \nabla 2 B \\ -\nabla T/T \end{array}\right),\label{sigma}
\end{equation}
where $J$'s are currents, with superscript $c$, $s$ and $e$ referring to charge, spin and energy, respectively; $\mu$, $B$ and $T$ denote chemical potential, magnetic field and temperature, respectively; $\sigma$'s denote various conductivities. In particular, the off-diagonal elements of the $\sigma$-matrix characterize couplings between different degrees of freedom.
When the gradient is sufficiently weak, the linear response theory introduced by Kubo often serves as an effective tool for studying the transport behavior \cite{Kubo:1957}.
Based on the Kubo formula, the transport coefficient is related to the time average of a two-point current correlator \cite{Kubo:book}.
For generic systems, we expect the conductivity to be finite in the late time due to the presence of scattering or Umklapp processes \cite{Callaway:1959,Hartnoll:2012,Maznev:2014,Rosch:2000,Nardis:2019}.
In contrast, there exists a class of integrable models that possesses infinitely many local conserved charges such that the conductivity does not relax thoroughly subject to the Mazur inequality \cite{Mazur:1969,Zotos:1997,Sirker:2011,Sirker:2020} and thus the conductivity $\sigma$ can be written as ${\rm Re}[\sigma(\omega)]=2\pi D \delta(\omega) +\sigma^{\rm reg}(\omega)$. Here, the regular part $\sigma^{\rm reg}$ characterizes the diffusion process, 
while the coefficient of the singular part, $D$, is the Drude weight (DW) \cite{Kohn:1964,Scalapino:1992}, which characterizes the dissipationless ballistic transport.
%

Based on the generalization of Kohn's formula to finite temperature \cite{Kohn:1964,Shastry:1990,Castella:1995,Zotos:1996,Narozhny:1998}, some analytic results are derived for the XXZ model \cite{Sirker:2011,Zotos:1990,Zotos:2017}. 
%
%
Starting from the dynamic assumption of local thermal equilibrium, the generalized hydrodynamics (GHD) approach is developed to study large space-time scale quantum dynamics in integrable one-dimensional (1D) systems
\cite{Castro:2016,Bertini:2016,Ilievski:2017}. 
The GHD predicts steady states through the Bethe-Boltzmann equations that govern the dynamics of quasiparticles with effective velocities
\cite{Ilievski:2017,Doyon:2017,Fava:2020,Bulchandani:2018}.
%
%
 In light of the GHD \cite{Nardis:2018,Nardis:2019,Panfil:2019,Doyon:2022,Ilievski:2018}, ballistic and diffusive dynamics can be successfully characterized
\cite{Ilievski:2018,Nardis:2019,Nardis:2018,Gopalakrishnan:2018,Gopalakrishnan:2019} and have been studied experimentally
\cite{Caux:2019,Schemmer:2019,Bastianello:2019,Malvania:2021,Ruggiero:2020}, see a recent review \cite{Guan:2022}.

However, the success of Kohn's formula and GHD is mainly based on numerical simulations on a few specific models \cite{Karrasch:2014,Rmer:1995,Ilievski:2017,Nozawa:2020,Nozawa:2021}. Analytic results of DW related to conserved charges other than the sole $U(1)$ charge are extremely difficult to obtain, and there exist only very few results in limiting situations such as zero temperature ground
state \cite{Shastry:1990,Rmer:1995}, half filled case \cite{Fujimoto:1998}, and infinite temperature limit \cite{Nozawa:2020}.
%
 It is therefore of fundamental interest to develop new methods to uncover universal spin and charge conductivities in strongly correlated electronic and atomic systems. 

In this Letter, by taking advantage of its quantum integrability, we rigorously calculate and derive the fundamental rules of DWs for the one-dimensional (1D) Fermi-Hubbard model with generic filling factor and arbitrary magnetic fields at low temperatures. This calculation bridges the Bethe ansatz (BA) or thermal Bethe ansatz (TBA) \cite{Lieb:1968,Lieb:2003,Ess05} and the GHD approaches. Our results establish general relations between the Luttinger parameters and the DWs, provide the microscopic origin of the (spin) Seebeck effects, and uncover counterintuitive spin-charge coupling and calorimetric effects. 
We further derive close forms of the DWs in quantum critical regimes, related to the Kardar-Parisi-Zhang (KPZ) dynamics identified by the DW curvature \cite{Fava:2020,Nardis:2020}. 
%
%
Based on recent experimental development of measuring DWs \cite{Tajik:2024}, we also propose a protocol to measure DWs in a 1D repulsive Hubbard model. 
%

{\em Charge and spin fluxes and DWs ---} By considering $U(1)\otimes U(1)$ symmetry \cite{Guan:1998} local gauge transformation on the fermionic operators with fluxes denoted by $\phi_{\uparrow},\phi_{\downarrow}$, the Hamiltonian of the 1D Hubbard model reads 
\begin{eqnarray}
	H&=&- \sum^L_{j=1, a=\uparrow,\downarrow}\left(\e^{\mi\phi_{a}/L}c_{j+1,a}^{\dagger} c_{j, a}+{\rm H.c.}\right)\no\\&&+u \sum_{j=1}^{L} (1-2n_{j \uparrow}) (1-2n_{j \downarrow})-\mu \hat{n} -2B\hat{S}^z,  \label{Ham}
\end{eqnarray}
where $c_{j,a}^\dagger$ ($c_{j,a}$) is the creation (annihilation) operator of an electron with
spin $a$ ($a=\uparrow$ or $\downarrow$) at site $j$ on a 1D lattice of length $L$. 
$\mu$ and $B$ are, respectively, the chemical potential and the magnetic field,
and $\hat{n}=n_{\uparrow}+n_{\downarrow},\,\hat{S}^z=(n_{\uparrow}-n_{\downarrow})/2$ are the density and magnetization operators, respectively. 
Hamiltonian (\ref{Ham}) is integrable under the $U(1)\otimes U(1)$ symmetry \cite{Guan:1998},  preserving density and magnetization. 
%
%
Following Refs.~\cite{Castella:1995,Rmer:1995,Urichuk:2022}, here we define general linear DWs as 
\begin{eqnarray}
	D^{\alpha \beta}&=&\frac{L}{2}\left\langle \frac{\partial^{2}E}{\partial \phi^{\alpha} \partial \phi^{\beta}}\right\rangle \bigg|_{\phi^{\alpha},\phi^{\beta}=0},
	\,\,\, \alpha, 
    \beta=c,s, 
	\label{Dcs0}
\end{eqnarray}
where $\phi^{c}=\left(\phi_{\uparrow}+\phi_{\downarrow}\right)/2, \,\phi^{s}=\left(\phi_{\uparrow}-\phi_{\downarrow}\right)/2$, and $\left\langle \cdots \right\rangle$ stands for thermal average. 
$D^{cc}$ and $D^{ss}$ denote the charge and spin DWs, respectively, while $D^{cs}$ and $D^{sc}$ are the cross DWs characterizing the coupling between the two degrees of freedom. By the definition of Eq.~(\ref{Dcs0}), we have $D^{cs}=D^{sc}$.

%

The key step to obtain the DWs is to determine the curvature of energy with respect to the fluxes $\phi^{c,s}$. 
%
%
Following the BA and TBA procedures with finite-size corrections \cite{Fujimoto:1998,Kohn:1964,Shastry:1990,Castella:1995,Zotos:1996,Narozhny:1998},  we expand the rapidities in powers of $1/L$: 
\begin{equation}
\theta_{\gamma} =\theta_{\gamma } ^{\infty}+\frac{x_{\gamma, 1}}{L}+\frac{x_{\gamma, 2}}{L^2}+O\left({1}/{L^3}\right).\label{x-alpha}
\end{equation}
Here, $\theta_{\gamma}$ presents the rapidities of $\gamma$-type quasiparticles with respect to the BA roots $k, \Lambda_n$ and $\Lambda_n^{'}$, representing quasiparticles of charge, length-$n$ spin string and $k$-$\Lambda$ string, respectively (see Supplemental Material (SM) \cite{JJL:SM} for details). 
The leading order $\theta_{\gamma } ^{\infty}$ obeys the standard BA and TBA equations \cite{Lieb:1968,Ess05}) which determine the energy density $e_0$ in the thermodynamic limit and are independent of the fluxes.
The total energy $E$ can also be expanded as 
\begin{equation}
\frac{E}{L}=e=e_0+\frac{E_1}{L}+\frac{E_2}{L^2}+\cdots.  \label{e-fsc}
\end{equation}
In the SM \cite{JJL:SM}, we prove that the first-order correction $E_1=0$, and the curvature of the second-order correction $E_2$ gives a general form of linear DWs
\begin{equation}
	D^{\alpha \beta}\!=\frac{1}{2T}\sum_{\gamma}\!\!\int \! d\theta_{\gamma}\rho_{\gamma}(1-n_{\gamma})v_{\gamma}^2 \! \prod_{\delta=\alpha, \beta}2\pi(\rho_{\gamma}+\rho_{\gamma}^h)\frac{dx_{\gamma,1}}{d\phi^{\delta}},\label{D-cssc}
\end{equation}
where $\rho_{\gamma}(\rho_{\gamma}^h)$ is the density of the particle (hole), $n_{\gamma}$ and $v_{\gamma}$ are, respectively, the occupation number and the effective velocity of type-$\gamma$ quasiparticle \cite{JJL:SM}.
Comparing (\ref{D-cssc}) with the GHD result \cite{Ilievski:2017,Doyon:2017,Fava:2020}, we further identify the dressed charges as
\begin{equation}
	{q_{\gamma}^{\alpha}}^{\dr}=\tau_\gamma 2\pi(\rho_{\gamma}+\rho_{\gamma}^h)\frac{dx_{\gamma,1}}{d\phi^{\alpha}},\label{q}
\end{equation}
$\tau_\gamma=1,1,-1$ for $\gamma=k,\, \Lambda_n,\, \Lambda_n^{'} $,  respectively. 
The dressed charges, playing the role analogous to mass in classical mechanics, are related to the bare charges $q_\gamma^\alpha$ as
\begin{equation}
{q_{\gamma}^{\alpha}}^{\dr}=\left(\textmd{I}-\bm{B}\right)_{\gamma \gamma^{\prime}}^{-1}*q_{\gamma^{\prime}}^{\alpha},\label{dr}
\end{equation}
where explicit expression of the matrix $\bm{B}$ that determines the dressing transformation is given  in SM \cite{JJL:SM},
and the bare charges $q_{\gamma}^{\alpha}$  for the charge and spin $U(1)$ charges   are  given by  
\begin{eqnarray} q_{k}^{c}&=&1,\;q_{\Lambda_n}^{c}=0,\;q_{\Lambda_n^{'}}^{c}=-2n_{\Lambda^{'}_n},\no\\
 q_{k}^{s}&=&1,\;q_{\Lambda_n}^{s}=-2n_{\Lambda_n},\;q_{\Lambda_n^{'}}^{s}=0,  \label{q0}
\end{eqnarray}
respectively.
Combining Eqs. (\ref{D-cssc}) and (\ref{dr}), thus we have a remarkable relationship between the DWs and the dressed charges:
\begin{equation}
	D^{\alpha \beta}\!=\frac{1}{2T}\sum_{\gamma}\!\!\int \! d\theta_{\gamma} \,\rho_{\gamma}(1-n_{\gamma})\,v_{\gamma}^2  \,{q_{\gamma}^{\alpha}}^{\dr} {q_{\gamma}^{\beta}}^{\dr}.\label{D-dr1}
\end{equation}

\begin{figure*}[ht] 
	\begin{center} 
		\includegraphics[width=1\linewidth]{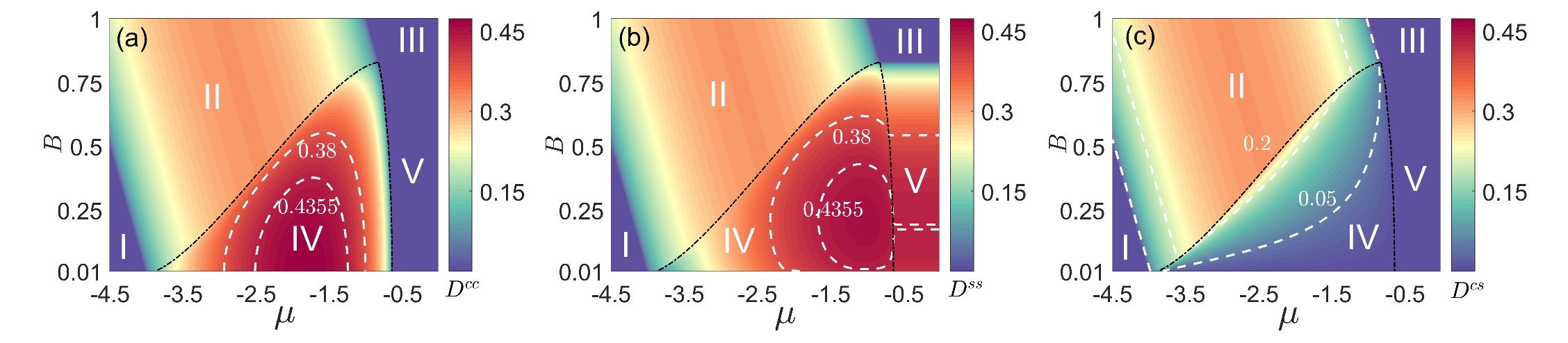} 
	\end{center}
\vspace{-15pt}  
	\caption{Full phase diagram represented by the contour plot of (a) charge (b) spin and (c) cross DWs at temperature $T=0.005$ and $u=1$. The dotted
		lines represent analytic solutions of BA equations obtained at
		zero temperature. The white dashed lines are the contour lines at $D^{cc},D^{ss}=0.38,0.4355$ and $D^{cs}=0.05,0.2$. Phases I to V correspond to vacuum, fully polarized partially filled, fully polarized half filled, partially polarized partially filled, and partially polarized half filled, respectively. The charge and cross DWs cannot distinguish phases III and V, which are the two Mott phases of charge insulator. For the two fully polarized phases (II and III), there is no distinction between spin and charge and, hence, all three DWs are the same.}      
	\label{Fig1:dcds}
\end{figure*}

In Fig.~\ref{Fig1:dcds}, we plot low temperature DWs in various phases in the parameter space spanned by $\mu$ and $B$. The charge and cross DWs vanish in Phases III and V, indicating the charge insulator nature of these two phases.  
The peak value for charge DW occurs within Phase IV, indicating maximum contributions from the spin-imbalanced states $\Lambda_1$ at low temperature. 
 %
%
%
%
In the Tomonaga-Luttinger liquid (TLL) regimes (Phases II and IV), we can safely ignore the gapped strings \cite{Luo:2023-PRB,Luo:2023}, and obtain universal DWs at low temperature as
	\begin{align}
		D^{\alpha \beta}=\!\!\!\sum_{\gamma=k,\Lambda}\! \frac{1}{2\pi}\!\!\left[{{q_{\gamma}^{\alpha}}^{\dr}{q_{\gamma}^{\beta}}^{\dr}}v_{\gamma} \!+\!\frac{\pi^2 T^2}{6}\frac{\partial^2({{q_{\gamma}^{\alpha}}^{\dr}{q_{\gamma}^{\beta}}^{\dr}}v_{\gamma})}{\partial (e_{\gamma}^{\ph})^2}\Big|_{e_{\gamma}^{\ph}=0} \right] ,\label{TLL}
\end{align}
where $e_{\gamma}^{\ph}$ are the physical energies of the quasiparticles \cite{JJL:SM}. 
This result shows that the DWs consist of separated contributions from the charge and spin sectors,  demonstrating  spin-charge intrinsic correlation \cite{Boll:2016,Hilker:2017,Vijayan:2020,Spar:2022} from a quantum transport perspective.

\begin{figure}[t] 
	\begin{center} 
		\includegraphics[width=0.8\linewidth]{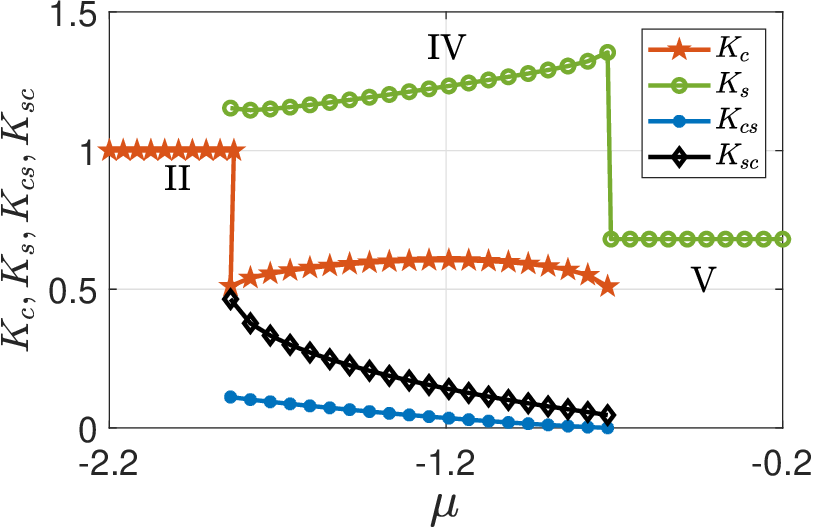} 
	\end{center}
	\vspace{-14pt}  
	\caption{Luttinger parameters vs chemical potential for $B=0.6$ and $u=1$ in the ground state.}         
	\label{Fig2:LP}
\end{figure}

{\em DWs and Luttinger parameters ---} In SM \cite{JJL:SM}, we prove that Eq.~(\ref{D-cssc}) establishes a general relation between DWs and Luttinger parameters in the low temperature regime for the Hubbard model with arbitrary external potentials.  The Luttinger parameters in terms of the dressed charges Eq.~(\ref{dr}) are given by 
\begin{eqnarray}
	K_c&=&\frac{{q_k^{c,\dr}}^2}{2},\, K_s=\frac{{q_{\Lambda}^{s,\dr}}^2}{2},\, 
	K_{cs}=\frac{{q_{\Lambda}^{c,\dr}}^2}{2},\, K_{sc}=\frac{{q_{k}^{s,\dr}}^2}{2},
	\no
\end{eqnarray}
where $K_c$, $K_s$ are the usual charge and spin Luttinger parameters, respectively, while we define $K_{cs},K_{sc}$ to be the cross Luttinger parameters characterizing the contributions of one degree of freedom to the other and capturing the spin-charge coupling effect. In Fig.~\ref{Fig2:LP}, we plot the Luttinger parameters as functions of $\mu$ for a fixed magnetic field $B$ and interaction strength $u$. The region we plot spans Phases II, IV and V. Phase II is a fully polarized phase and hence represents non-interacting spinless fermions. It possesses only charge degrees of freedom with $K_c=1$. Phase V is a charge insulating phase and possesses only spin degrees of freedom described by an isotropic spin chain with $1/2\le K_s<1$, where the equal sign occurs for zero magnetic field.  Both degrees of freedom are present in Phase IV, where $K_c<1<K_s$ due to the repulsive interaction. $K_c$ and $K_s$ exhibit sharp jumps at phase boundaries. For a zero magnetic field, the cross Luttinger parameters $K_{cs, sc}$ vanish as a result of the spin rotation symmetry \cite{JJL:SM}.

With these definitions of Luttinger parameters and taking the zero temperature limit of Eq.~(\ref{TLL}), we immediately have
\begin{eqnarray}
		D^{cc}&=&\frac{K_cv_c}{\pi}+\frac{K_{cs}v_s}{\pi},  \quad D^{ss}=\frac{K_sv_s}{\pi}+\frac{K_{sc}v_c}{\pi}, \no\\
		D^{cs}&=&D^{sc}=\frac{\sqrt{K_cK_{sc}}\,v_c}{\pi}+\frac{\sqrt{K_sK_{cs}}\,v_s}{\pi},\label{dc-ds-bm}	
\end{eqnarray}
which establishes an insightful relation between the Luttinger parameters and the DWs.
Here, we have rewritten the charge and spin velocities $v_k$, $v_\Lambda$ using the more familiar notation $v_c$, $v_s$, respectively. For zero magnetic field, the cross Luttinger parameters, and hence the cross DW, vanish, and we recover the previous bosonization result: $D^{cc}=(K_cv_c)/\pi$, $D^{ss}=(K_sv_s)/\pi$ \cite{Giamarchi:1997,Giamarchi-book}.
\begin{figure}[!htb] 
	\begin{center} 
		\includegraphics[width=0.9\linewidth]{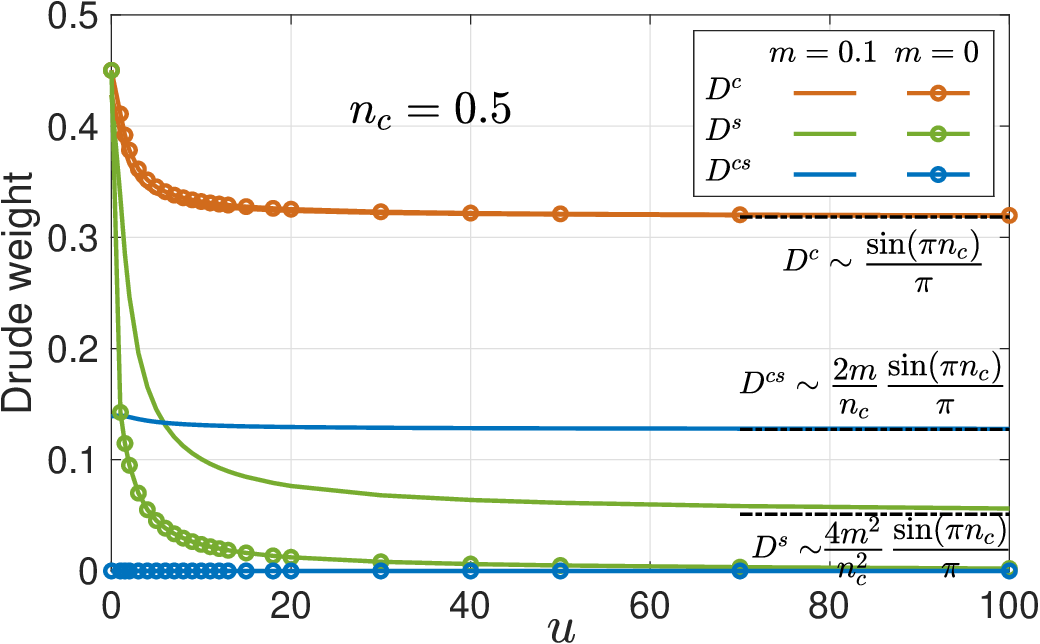} 
	\end{center} 
\vspace{-14pt} 
	\caption{{Ground-state DWs as functions of interaction strength $u$ at the filling factor   $n_c=0.5$ and for two different magnetization densities $m=0$ and $m=0.1$. All lines are analytic results of Eqs.~(\ref{dc-ds-bm}).}}           
	\label{Fig3:dw}
\end{figure}

The lines in Fig.~\ref{Fig3:dw} represent the DWs for the ground state as functions of the interaction strength $u$ based on Eqs.~(\ref{dc-ds-bm}) for fixed  filling factor $n_c=(n_\uparrow+n_\downarrow)/L$, and magnetization densities $m=(n_\uparrow-n_\downarrow)/(2L)$.
Here, we see that both the spin and charge DWs decrease as $u$ increases, while the cross DW is not sensitive to $u$. All DWs converge to their respective asymptotic values in the large $u$ limit, which can be obtained as
\begin{eqnarray}
    D^{cc}&=&\frac{\sin(\pi n_c)}{\pi},\quad D^{ss}=\frac{4m^2}{\pi n_c^2}\sin(\pi n_c),\, \no \\
    D^{cs}&=&D^{sc}=\frac{2m}{\pi n_c}\sin(\pi n_c). \label{largeu}
\end{eqnarray}
We observe that the asymptotic values of the charge DW are independent of spin imbalance $m$, while those of the spin and cross DWs depend on $m$ in a quadratic and linear manner, respectively. 
In this limit, for a finite magnetization in phase IV, even though $v_s=0$, the spin and cross DWs can still exist, showing a counterintuitive behavior of transport. 
They have finite values due to the coupling between the spin and charge degrees of freedom, manifested by the finite cross Luttinger parameters.

{\em Spin and charge Seebeck effect ---} 
The Seebeck effect (SE) refers to a thermoelectric phenomenon in which a temperature gradient induces an electric current \cite{Geballe1955,Herwaarden1986,Goldsmid2010}. 
More recently, this concept has been extended to the spin sector in which a spin current is induced by the temperature gradient. Spin SE is expected to be vital for the development of spin caloritronics in future quantum devices \cite{adachi2013theory,Uchida2008,Jaworski2010,Barfknecht2021}.
Despite significant interest, rigorous calculations of the transport coefficients --- specifically $D^{se}$ and $D^{ce}$ --- associated with (spin) SE have not been carried out so far. Our approach can be straightforwardly extended to calculate these coefficients.

In terms of GHD, the cross DW $D^{se}=D^{es}$ and $D^{ce}=D^{ec}$ can still be universally expressed in the same form as Eqs.~(\ref{D-dr1}) and (\ref{TLL}), with one of the dressed charges replaced by the dressed energy ${q_{\gamma}^{e}}^{\dr}$, whose explicit expressions  can be found in SM \cite{JJL:SM}.

\begin{figure}[!htb] 
	\begin{center} 
		\includegraphics[width=1\linewidth]{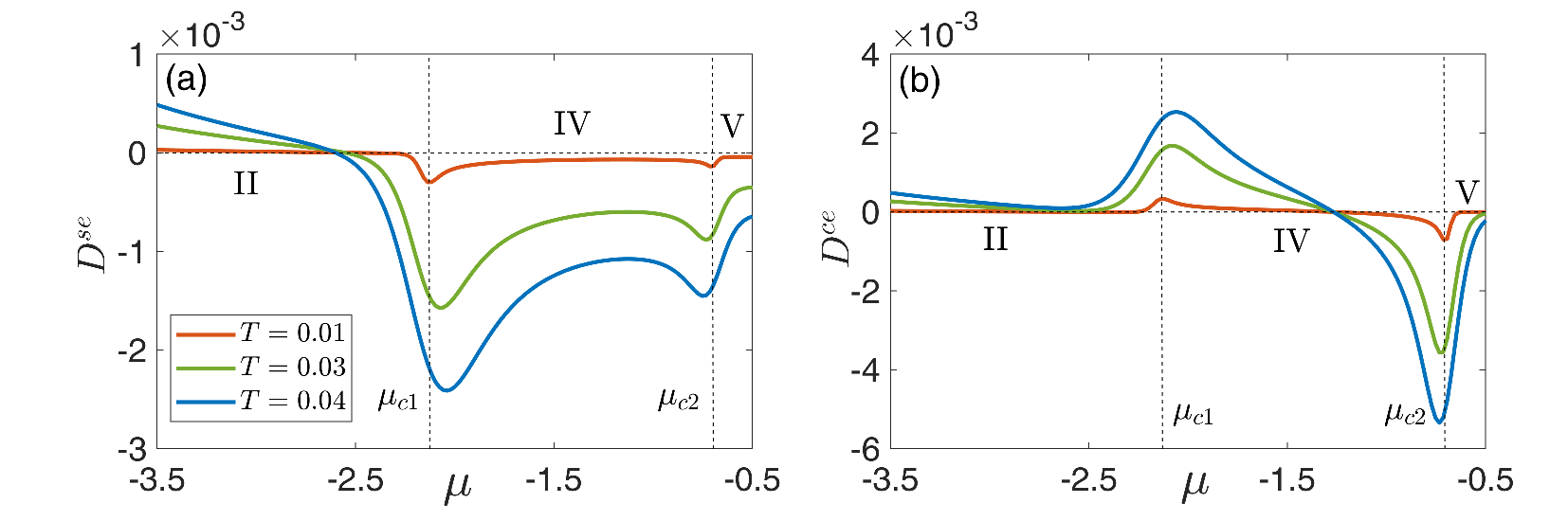} 
	\end{center}
	\vspace{-14pt}  
	\caption{{$D^{se}$ (a) and $D^{ce}$ (b) as functions of chemical potential $\mu$ for different temperatures $T$ at $B=0.5,u=1$. The critical fields $\mu_{c1},\mu_{c2}$, indicated as vertical black dashed lines, are boundaries for II-IV and IV-V, respectively. }}    
	\label{Fig4}
\end{figure}

At zero temperature, the dressed energies ${q_{k}^{e}}^{\dr}$ and ${q_{\Lambda}^{e}}^{\dr}$ both vanish at their respective Fermi points. This leads to $D^{se}=D^{ce}=0$, indicating a vanishing spin and charge Seebeck effect.
We further prove that $D^{se}$ and $D^{ce}$ share the same scaling behavior with entropy in the quantum critical regime \cite{JJL:SM}, leading to the enhancement  of spin and charge currents near phase boundaries. To demonstrate this, we plot in Fig.~\ref{Fig4} $D^{se}$ and $D^{ce}$ as functions of the chemical potential $\mu$ for different temperatures. The parameter range traverses Phases II, IV and V.
One can observe from Fig.~\ref{Fig4} the clear enhancement of the magnitude of $D^{se}$ and $D^{ce}$ around the critical points. As in the previous case, $D^{se}$ and $D^{ce}$ receive a separate contribution from the spin and charge degrees of freedom. These separate contributions,  leading to the inversion of the SEs,  are discussed in the SM \cite{JJL:SM}.

%

%

{\em DWs at quantum criticality ---}
Previous studies have shown that the charge DW is useful in probing the Mott phase transition \cite{Giamarchi:1997,Moore:2007,Koretsune:2007}. 
 In Fig.~\ref{Fig1:dcds}, we show that spin and charge DWs can be used to distinguish all phases. 
In the SM \cite{JJL:SM}, we present universal scaling behavior for various phase transitions at low temperature. 
%
To give one example, for the phase transition II-IV , we observe that at low temperature the spin and charge DWs consist of separate contributions from the charge and spin sectors with the following universal scaling forms
\begin{align}
    D_{\Lambda}^{\alpha\beta}=&\frac{2}{\sigma(0)}\left(\frac{{q_{\Lambda}^{\alpha}}^{\dr}(0)}{2\pi}\right)\left(\frac{{q_{\Lambda}^{\beta}}^{\dr}(0)}{2\pi}\right)\left(\frac{\varepsilon_1^{\prime \prime}(0)}{2}\right)f_{\frac{1}{2}},\no\\
	D_{k}^{\alpha\beta}=&D_{(0)}\left[1-f_{\frac{1}{2}}\sum_{e=\alpha, \beta}\frac{\frac{1}{2}\sigma(0){{q_{\Lambda}^{e}}^{\dr}}(0)^{\prime}-\sigma(0)^{\prime}{q_{\Lambda}^{e}}^{\dr}(0)}{\rho^0{{q_{k}^{e}}^{\dr}}^0}\right]\no\\&+\frac{\pi^2T^2}{6}\frac{\partial^2}{\partial \kappa^2}D_{(0)}|_{\kappa=0},
    \label{dwqc}
\end{align}
where $\sigma(0)$ $({q_{\Lambda}^{e}}^{\dr}(0))$ denotes the spin density (spin component of dressed charges) at $\Lambda=0$;
$\rho^0,\,{{q_{k}^{e}}^{\dr}}^0$ denote their values at zero temperature; the primes represent the derivatives with respect to the relevant parameters;
 %
  $D_{(0)}$ denotes the background contribution from the ground state, and $\varepsilon_1$ gives the physical energy of the length-1 $\Lambda$ string. We also denote 
\begin{eqnarray}
	f_{\frac{1}{2}}&\equiv&-\frac{1}{2}\left(\frac{\varepsilon_1^{\prime \prime}(0)}{2}\right)^{-\frac{1}{2}}\pi^{\frac{1}{2}}T^{\frac{1}{2}}\Li_{\frac{1}{2}}\left(-\e^{\frac{-\varepsilon_1(0)}{T}}\right),\label{li}
\end{eqnarray}
where $\mathrm{Li}_n$ denotes the polylog functions, $\varepsilon_1(0)$ measures the distance from the critical point
\cite{Luo:2023}. 
From Eq.~(\ref{li}), we observe that the scaling laws of the DW are consistent with those of the first-order thermodynamic quantities with the dynamic exponent $z=2$ and the correlation exponent $\nu=1/2$. 
%
 Such a scaling behavior reveals a universal law with respect to temperature, which can be directly applied to other integrable systems.
%

%

%
%

%
 
%
{\em Experimental protocol ---} Motivated by the recent experiment \cite{Tajik:2024} in which the charge DW of a 1D spinless Bose gas is measured, here we propose an experimental protocol to measure the DWs studied in this work.
We consider a system where repulsive fermions are trapped in an optical lattice with an additional box potential.
Then a weak linear charge or spin gradient is suddenly turned on and will induce charge/spin transport with a current displaying time-asymptotic growth in the long time limit. By monitoring the currents, the DWs can be extracted. 
We directly simulate this protocol numerically with details given in the SM \cite{JJL:SM}. We show that the numerical results are in good agreement with the analytic ones presented in the main text.

{\em Summary and outlook ---} Using the generalized Kohn's formula, we have analytically  obtained the close forms of various DWs in the 1D Hubbard model with arbitrary magnetic fields and chemical potentials. Our work builds up a direct connection to the GHD, and reveals the fundamental rules of ballistic transport of quasiparticles in terms of Luttinger parameters. 
We have shown that in the magnetized phases with spin imbalance, the transport dynamics exhibits a spin-charge coupling effect that can be captured by the cross DWs $D^{cs}$ and $D^{se}$. Our calculation builds a microscopic origin of the (spin) Seebeck effect in which a temperature gradient induces a charge/spin current. We show that the associated cross DWs $D^{ce}$ and $D^{se}$ obey the same quantum critical scaling law as entropy across various phase boundaries. Our work not only provides deeper insights into quantum transport of quantum many-body systems, but can also find potential applications in quantum spintronic and calorimetric devices. 
Moreover, our methods can be extended to the study of spin and charge diffusion in ultracold atoms \cite{Matthew:2019,Bardon:2014} and persistent current in atomtronics \cite{Amico:Atomtronic}. We plan to carry out such studies in the near future.

\begin{acknowledgments}
{\em  Acknowledgments--}
We thank Prof. Shizhong Zhang for valuable comments and suggestions.
X.W.G acknowledges support from the NSFC key grants No. 92365202, No. 12134015, and the National Key R\&D Program of China under Grant No. 2022YFA1404102. He is also partially supported by the Innovation Program for Quantum Science and Technology 2021ZD0302000. H.P. acknowledges support from the U.S. NSF and the Welch Foundation (Grant No. C-1669). S.B. acknowledges support from the U.S. NSF through Grant Number PHY-2409311.\\
 {\em   Note added--}The recent paper arXiv: 2506.00686 by F. G\"{o}hmann, A. Kl\"{u}mper and K. K. Kozlowski reports   the particle transport in the mono-atomic Bose gas, showing a consistency with the charge Drude weight of sole charge degree of freedom $D^c =K_c v_c /\pi $  given in Eq. (12).  
\end{acknowledgments}

\clearpage\newpage
\setcounter{figure}{0}
\setcounter{table}{0}
\setcounter{equation}{0}
\def\thefigure{S\arabic{figure}}
\def\thetable{S\arabic{table}}
\def\theequation{S\arabic{equation}}
\setcounter{page}{1}
\pagestyle{plain}

\begin{widetext}

\section*{Quantum transport in 1D  Hubbard model: Drude weights and Seebeck effect}
\begin{center}
{Jia-Jia Luo, Sagarika Basak, Han Pu, Xi-Wen Guan}
\end{center}

\section{Drude weight from finite size correction}

\subsection{Bethe ansatz equations with twisted boundary condition}
The 1D one-band Hubbard model possesses $U(1)\otimes U(1)$ hidden gauge symmetry \cite{Guan:1998}. By virtue of this symmetry upon Yang-Baxter relation, it can be proven that this model is integrable. One case of this transformation is the Hubbard model with twisted boundary condition, namely 
$c_{L+1,a}^{\dagger}=\e^{\mi\phi_{a}}c_{1,a}^{\dagger}$ with fluxes denoted by $\phi_{\uparrow},\phi_{\downarrow}$. After this operation,
the Hamiltonian takes the form
\begin{equation}
	H=-\sum_{j=1}^{L} \sum_{a=\uparrow \downarrow}\left(\e^{\mi\phi_{a}/L}c_{j+1,a}^{\dagger} c_{j, a}+{\rm H.c.}\right)+u \sum_{j=1}^{L} (1-2n_{j \uparrow}) (1-2n_{j \downarrow}),
\end{equation}
or a different but equivalent representation
\begin{equation}
	H=-\sum_{j=1}^{L-1} \sum_{a=\uparrow \downarrow}\left(c_{j+1,a}^{\dagger} c_{j, a}+\rm{H.c.}\right)-\sum_{a=\uparrow \downarrow}\left(\e^{\mi\phi_{a}}c_{1,a}^{\dagger} c_{L, a}+\rm{H.c.}\right)+u \sum_{j=1}^{L} (1-2n_{j \uparrow}) (1-2n_{j \downarrow}).
\end{equation}
Then extra phase factors will appear within Bethe ansatz (BA) equations
\begin{equation}
\begin{aligned}
\e^{\mi k_jL}=\e^{\mi\phi_{\uparrow}}\prod_{L=1}^{M}\frac{\lambda_l-\sin k_j-\mi u}{\lambda_l-\sin k_j+\mi u},j=1,\cdots,N,\\
\e^{\mi\left(\phi_{\uparrow}-\phi_{\downarrow}\right)}\prod_{j=1}^{N}\frac{\lambda_l-\sin k_j-\mi u}{\lambda_l-\sin k_j+\mi u}=-\prod_{m=1}^{M}\frac{\lambda_l-\lambda_m-2\mi u}{\lambda_l-\lambda_m+2\mi u},l=1,\cdots,M	,
\end{aligned}
\end{equation}
where $N$ is the total particle number, $M$ is spin-down electron number.  Inserting rapidities expressions of $\Lambda,k-\Lambda$ strings into above BA equations, the \textit{Takahashi's equations} are reconstructed as
\begin{equation}
\begin{aligned}
k_{j} L=&\phi_{\uparrow}+2 \pi I_{j}-\sum_{n=1}^{\infty} \sum_{\alpha=1}^{M_{n}}
 \theta\left(\frac{\sin k_{j}-\Lambda_{\alpha}^{n}}{n u}\right)-\sum_{n=1}^{\infty} \sum_{\alpha=1}^{M_{n}^{\prime}} \theta\left(\frac{\sin k_{j}-\Lambda_{\alpha}^{\prime n}}{n u}\right), \\
\sum_{j=1}^{N-2 M^{\prime}} \theta\left(\frac{\Lambda_{\alpha}^{n}-\sin k_{j}}{n u}\right)=&-n\left(\phi_{\uparrow}-\phi_{\downarrow}\right)+2 \pi J_{\alpha}^{n}+\sum_{m=1}^{\infty} \sum_{\beta=1}^{M_{m}} \Theta_{n m}\left(\frac{\Lambda_{\alpha}^{n}-\Lambda_{\beta}^{m}}{u}\right), \\
2 L \operatorname{Re}\left[\arcsin \left(\Lambda_{\alpha}^{\prime n}+n i u\right)\right]=&n\left(\phi_{\uparrow}-\phi_{\downarrow}\right)-2n\phi_{\uparrow}+2 \pi J_{\alpha}^{\prime n}+\sum_{j=1}^{N-2 M^{\prime}} \theta\left(\frac{\Lambda_{\alpha}^{\prime n}-\sin k_{j}}{n u}\right)\\&+\sum_{m=1}^{\infty} \sum_{\beta=1}^{M_{m}^{\prime}} \Theta_{n m}\left(\frac{\Lambda_{\alpha}^{n}-\Lambda_{\beta}^{\prime m}}{u}\right),
\end{aligned}
\end{equation}
where $\Lambda_{\alpha}^{n},\Lambda_{\alpha}^{\prime n}$ are real parts of relevant strings. We can conduct finite-size corrections of the rapidities as
\begin{equation}
\begin{aligned}
k_j=&k_j^{\infty}+\frac{x_1}{L}+\frac{x_2}{L^2}+O\left(\frac{1}{L^3}\right),\\
\Lambda_{\alpha}^n=&\Lambda_{\alpha}^{n\infty}+\frac{y_{1n}}{L}+\frac{y_{2n}}{L^2}+O\left(\frac{1}{L^3}\right),\\
\Lambda_{\alpha}^{\prime n}=&\Lambda_{\alpha}^{\prime n\infty}+\frac{z_{1n}}{L}+\frac{z_{2n}}{L^2}+O\left(\frac{1}{L^3}\right),\label{fsc}
\end{aligned}
\end{equation}
where $k_j^{\infty},\Lambda_{\alpha}^{n\infty},\Lambda_{\alpha}^{\prime n\infty}$ represent relevant rapidities in the thermodynamic limit. Substituting above rapidities expressions into BA equations and using Euler-McLaurin summation formula, we derive the density equations with the first three orders of $(1/L)$ as follows:

$O(1):$
\begin{equation}
\begin{aligned}
\rho^p(k)+\rho^h(k)=&\frac{1}{2\pi}+\cos k \sum_{n=1}^{\infty} \int_{-\infty}^{\infty} \d \Lambda a_{n}(\sin k-\Lambda)\left(\sigma_{n}^{\prime p}(\Lambda)+\sigma_{n}^p(\Lambda)\right), \\
\sigma_{n}^h(\Lambda) =& -\sum_{m=1}^{\infty} A_{n m} * \sigma_{m}^p(\Lambda) +\int_{-\pi}^{\pi} \d k a_{n}(\sin k-\Lambda) \rho^p(k),\\
\sigma_{n}^{\prime h}(\Lambda) =&\frac{1}{\pi} \text{Re} \frac{1}{\sqrt{1-(\Lambda-\mi n u)^{2}}} -\sum_{m=1}^{\infty} A_{n m} * \sigma_{m}^{\prime p}(\Lambda) -\int_{-\pi}^{\pi} \d k a_{n}(\sin k-\Lambda) \rho^p(k),\label{den0}
\end{aligned}
\end{equation}

$O(1/L):$
\begin{equation}
\begin{aligned}
	\left(\rho^p(k)+\rho^h(k)\right)x_1=&\frac{\phi_{\uparrow}}{2\pi}+\sum_{n=1}^{\infty} \int_{-\infty}^{\infty} \d \Lambda a_{n}(\sin k-\Lambda)\left(z_{1n}\sigma_{n}^{\prime  p}(\Lambda)+y_{1n}\sigma_{n}^p(\Lambda)\right), \\
	\sigma_{n}^h(\Lambda)y_{1n} =& -\frac{n\left(\phi_{\uparrow}-\phi_{\downarrow}\right)}{2\pi}-\sum_{m=1}^{\infty} A_{n m} * y_{1m}\sigma_{m}^p(\Lambda) +\int_{-\pi}^{\pi} \d k a_{n}(\sin k-\Lambda) x_1\cos k\rho^p(k),\\
	\sigma_{n}^{\prime h}(\Lambda)z_{1n} =&-\frac{n\phi_{\uparrow}}{\pi}+\frac{n\left(\phi_{\uparrow}-\phi_{\downarrow}\right)}{2\pi} -\sum_{m=1}^{\infty} A_{n m} *z_{1m} \sigma_{m}^{\prime p}(\Lambda) -\int_{-\pi}^{\pi} \d k a_{n}(\sin k-\Lambda) x_1\cos k\rho^p(k),\label{den1}
\end{aligned}
\end{equation}	

$O(1/L^2):$
\begin{equation}
\begin{aligned}
 \left(\rho^p(k)+\rho^h(k)\right)x_2=&\sum_{n=1}^{\infty} \int_{-\infty}^{\infty} \d \Lambda a_{n}(\sin k-\Lambda)
 \left(z_{2n}\sigma_{n}^{\prime  p}(\Lambda)+y_{2n}\sigma_{n}^p(\Lambda)\right)+\frac{1}{2}\frac{\partial}{\partial k}\left[(\rho^p+\rho^h)x_1^2\right]\\&+\frac{1}{2}\sum_{n=1}^{\infty} \int_{-\infty}^{\infty} \d \Lambda \frac{\partial}{\partial \Lambda}[a_{n}(\sin k-\Lambda)]\left(z_{1n}^2\sigma_{n}^{\prime  p}(\Lambda)+y_{1n}^2\sigma_{n}^p(\Lambda)\right),
 \\
 \left(\sigma_{n}^p(\Lambda)+\sigma_{n}^h(\Lambda)\right)y_{2n}=&\frac{1}{2}\frac{\partial}{\partial \Lambda}\left[(\sigma_{n}^p+\sigma_{n}^h)y_{1n}^2\right]+\int_{-\pi}^{\pi} \d k a_{n}(\Lambda-\sin k) x_2\cos k\rho^p(k)\\&-\sum_{m=1}^{\infty}\int_{-\infty}^{\infty} \d \Lambda^{\prime}\frac{1}{2\pi}\frac{\partial}{\partial \Lambda}\Theta_{nm}\left(\frac{\Lambda-\Lambda^{\prime}}{u}\right)y_{2m}\sigma_m^p\\&-\frac{1}{4\pi}\int_{-\pi}^{\pi} \d k\frac{\partial^2}{\partial k^2}\theta\left(\frac{\Lambda-\sin k}{nu}\right)x_1^2\rho(k)\\&+\sum_{m=1}^{\infty}\frac{1}{4\pi}\int_{-\infty}^{\infty} \d \Lambda^{\prime}\frac{\partial^2}{\partial \Lambda^2} \Theta_{nm}\left(\frac{\Lambda-\Lambda^{\prime}}{u}\right) y_{1m}^2\sigma_m^p\\&+\sum_{m=1}^{\infty}\lim_{\Lambda_0\rightarrow\infty}\frac{1}{48\pi u}\left[\frac{\Theta_{nm}^{\prime}((\Lambda-\Lambda_0)/u)}{\sigma_{m}^p(\Lambda_0)+\sigma_{m}^h(\Lambda_0)}-\frac{\Theta_{nm}^{\prime}((\Lambda+\Lambda_0)/u)}{\sigma_{m}^p(-\Lambda_0)+\sigma_{m}^h(-\Lambda_0)}\right],
 \\
 \left(\sigma_{n}^{\prime p}(\Lambda)+\sigma_{n}^{\prime h}(\Lambda)\right)z_{2n} =&\frac{1}{2}\frac{\partial}{\partial \Lambda}\left[(\sigma_{n}^{\prime p}+\sigma_{n}^{\prime h})z_{1n}^2\right]-\int_{-\pi}^{\pi} \d k a_{n}(\Lambda-\sin k) x_2\cos k\rho^p(k)\\&-\sum_{m=1}^{\infty}\int_{-\infty}^{\infty} \d \Lambda^{\prime}\frac{1}{2\pi}\frac{\partial}{\partial \Lambda}\Theta_{nm}\left(\frac{\Lambda-\Lambda^{\prime}}{u}\right)z_{2m}\sigma_m^{\prime p}\\&+\frac{1}{4\pi}\int_{-\pi}^{\pi} \d k\frac{\partial^2}{\partial k^2}\theta\left(\frac{\Lambda-\sin k}{nu}\right)x_1^2\rho(k)\\&+\sum_{m=1}^{\infty}\frac{1}{4\pi}\int_{-\infty}^{\infty} \d \Lambda^{\prime}\frac{\partial^2}{\partial \Lambda^2} \Theta_{nm}\left(\frac{\Lambda-\Lambda^{\prime}}{u}\right) z_{1m}^2\sigma_m^{\prime p}\\&+\sum_{m=1}^{\infty}\lim_{\Lambda_0\rightarrow\infty}\frac{1}{48\pi u}\left[\frac{\Theta_{nm}^{\prime}((\Lambda-\Lambda_0)/u)}{\sigma_{m}^{\prime p}(\Lambda_0)+\sigma_{m}^{\prime h}(\Lambda_0)}-\frac{\Theta_{nm}^{\prime}((\Lambda+\Lambda_0)/u)}{\sigma_{m}^{\prime p}(-\Lambda_0)+\sigma_{m}^{\prime h}(-\Lambda_0)}\right]\label{den2},
\end{aligned}
\end{equation}
where the expressions of order $O(1)$ gives rise to the standard BA equations. Thus the other two expressions are important.

\subsection{Expressions of linear Drude weight}	
In a finite size system, the dependence of total energy $E$ on the length of the system is 
\begin{equation}
\frac{E}{L}=e=E_0+\frac{E_1}{L}+\frac{E_2}{L^2}+O(\frac{1}{L^3})\label{e-fsc},
\end{equation}
where $E$ is total energy, $E/L=e$ denotes energy density, and $E_1(E_2)$ corresponds to the first (second)-order correction. In the system under study, there are two distinct U(1) symmetries, associated with the conservation of charge $\hat{N}$ and magnetization $\hat{S}^z$, respectively.
Charge Drude weight (DW) $D^{cc}$ characterizes the flow of total particles or mass, while spin DW $D^{ss}$ measures the difference between the flow of spin-up and -down particles. The former requires $\phi_{\uparrow}=\phi_{\downarrow}$ while the latter requires $\phi_{\uparrow}=-\phi_{\downarrow}$. Additionally, we can also consider the effect of charge (spin) fluctuations on spin (charge) transport. They are characterized by the cross DWs $D^{sc}$ and $D^{cs}$. Based on this analysis and the size dependence of energy spectra, the charge, spin and cross DWs can be derived through the derivatives of energy density $e$ w.r.t. fluxes
\begin{eqnarray}
D^{cc}&=&\frac{L}{2}\left\langle\frac{\partial^2E}{\partial {\phi^c}^2}\right\rangle \bigg|_{\phi^c=0}\label{Dc0},\\
D^{ss}&=&\frac{L}{2}\left\langle\frac{\partial^2E}{\partial {\phi^s}^2}\right\rangle \bigg|_{\phi^s=0}\label{Ds0},\\
D^{cs}&=&\frac{L}{2}\left\langle\frac{\partial^2E}{\partial {\phi^c}\partial {\phi^s}}\right\rangle \bigg|_{\phi^c,\phi^s=0}\label{cs0},\\
D^{sc}&=&\frac{L}{2}\left\langle\frac{\partial^2E}{\partial {\phi^s}\partial {\phi^c}}\right\rangle \bigg|_{\phi^s,\phi^c=0}\label{sc0},
\end{eqnarray}
where $\phi^{c}=\left(\phi_{\uparrow}+\phi_{\downarrow}\right)/2, \phi^{s}=\left(\phi_{\uparrow}-\phi_{\downarrow}\right)/2$. The total energy of repulsive Hubbard model is
\begin{eqnarray}
\frac{E}{L}&=&\frac{1}{L}\sum_{j=1}^{M_e}(-2\cos k_j -\mu-2u-B)+\frac{1}{L}\sum_{n=1}^{\infty}
\sum_{\alpha=1}^{M_n^{\prime}}\left(4\textmd{Re} \sqrt{1-(\Lambda_{\alpha}^n-\mathrm{i} n u)^{2}}-2 n \mu-4 n u\right)\nonumber\\&&+\frac{1}{L}\sum_{n=1}^{\infty}
\sum_{\alpha=1}^{M_n}2nB+u\nonumber\\&=&\frac{1}{L}\sum_{j=1}^{M_e}\left[-2\cos k_j^{\infty} -\mu-2u-B+2\sin k_j^{\infty}\left(\frac{x_1}{L}+\frac{x_2}{L^2}\right)+2\cos k_j^{\infty} \frac{x_1^2}{2L^2}\right]\nonumber\\&&+\frac{1}{L}\sum_{n=1}^{\infty}
\sum_{\alpha=1}^{M_n^{\prime}}\left[D+D^{\prime} \left(\frac{z_{1n}}{L}+\frac{z_{2n}}{L^2}\right)+\frac{1}{2}D^{\prime\prime}\frac{z_{1n}^2}{L^2}-2n\mu\right]+\frac{1}{L}\sum_{n=1}^{\infty}
\sum_{\alpha=1}^{M_n}2nB+u,
\end{eqnarray}	
where $D\equiv 4\textmd{Re} \sqrt{1-(\Lambda_{\alpha}^n-\mathrm{i} n u)^{2}}-4 n u$,  $D^{\prime} (D^{\prime\prime})$ denotes the first- (second-)order derivative of $D$ w.r.t. $\Lambda_{\alpha}^n$, and $k_j^{\infty}$ is the quasimomentum in the limit of $L\rightarrow \infty$ (For brevity in the following discussion, we omit the superscript of quasimomentum and use $k_j$ instead of $k_j^{\infty}$.). Based on eq.(\ref{e-fsc}), we can separate the relevant contributions of each order. Thus the explicit expressions of $E_0,E_1,E_2$ are 
\begin{eqnarray}
E_0&=&\int_{-\pi}^{\pi} \d k(-2\cos k-\mu-2u-B)\rho^p(k)+\sum_{n=1}^{\infty}\int_{-\infty}^{\infty} \d \Lambda 2nB\sigma_{n}^p(\Lambda)+u\nonumber\\&&+\sum_{n=1}^{\infty}\int_{-\infty}^{\infty} \d \Lambda\left(4\textmd{Re} \sqrt{1-(\Lambda-\mathrm{i} n u)^{2}}-2 n \mu-4 n u\right)\sigma_{n}^{\prime p}(\Lambda)\label{E0},\\
E_1&=&\int_{-\pi}^{\pi} \d k 2\sin kx_1\rho^p(k)+\sum_{n=1}^{\infty}\int_{-\infty}^{\infty} \d \Lambda D^{\prime}z_{1n}\sigma_{n}^{\prime p}(\Lambda)\label{E1},\\
E_2&=&\int_{-\pi}^{\pi} \d k\left(2\sin k x_2+\cos kx_1^2\right)\rho^p(k)+\sum_{n=1}^{\infty}\int_{-\infty}^{\infty} \d \Lambda\left(D^{\prime}z_{2n}+D^{\prime\prime}\frac{z_{1n}^2}{2}\right)\sigma_{n}^{\prime p}(\Lambda)\label{E2}.
\end{eqnarray}
From above we can easily tell that $E_0$ is independent of $\phi$ and hence makes no contribution to the DWs, which are completely determined by the next two orders $E_1$ and $E_2$, whose $\phi$-dependence come from the terms $x_1(k),\,x_2(k),\,z_{1n}(\Lambda),\,z_{2n}(\Lambda)$. 

\subsubsection{Evalulation of $E_1$}
First, we show that $E_1=0$ and the proof is as follows. In terms of the relations between densities and dressed energies, namely $\rho^h/{\rho^p}=\exp(\kappa/T),\sigma_n^h/{\sigma_n^p}=\exp(\varepsilon_n/T),\sigma_n^{\prime h}/{\sigma_n^{\prime p}}=\exp(\varepsilon_n^{\prime}/T)$, and the definitions of occupation number
\begin{eqnarray}
n_k&=&\frac{\rho^p}{\rho^p+\rho^h}=\frac{1}{1+\exp(\kappa/T)},\\
n_m&=&\frac{\sigma_m^p}{\sigma_m^p+\sigma_m^h}=\frac{1}{1+\exp(\varepsilon_m/T)},\\
n_m^{\prime}&=&\frac{\sigma_m^{\prime p}}{\sigma_m^{\prime p}+\sigma_m^{\prime h}}=\frac{1}{1+\exp(\varepsilon_m^{\prime}/T)},
\end{eqnarray}
we rewrite eq.(\ref{den1}) as
\begin{eqnarray}
x_1\rho^p(k)&=&\frac{n_k\phi_{\uparrow}}{2\pi}+\sum_{n=1}^{\infty} \int_{-\infty}^{\infty} \d \Lambda n_ka_{n}(\sin k-\Lambda)\left(y_{1n}\sigma_{n}^p(\Lambda)+z_{1n}\sigma_{n}^{\prime  p}(\Lambda)\right) \label{den1m-1},\\
y_{1n}\sigma_{n}^p(\Lambda) &=& -n_n\frac{n\left(\phi_{\uparrow}-\phi_{\downarrow}\right)}{2\pi} +\int_{-\pi}^{\pi} \d k n_n\cos ka_{n}(\sin k-\Lambda) x_1\rho^p(k)\nonumber\\&&-\sum_{m=1}^{\infty}\int_{-\infty}^{\infty} \d \Lambda^{\prime}n_n\frac{1}{2\pi}\frac{\partial}{\partial \Lambda}\Theta_{nm}\left(\frac{\Lambda-\Lambda^{\prime}}{u}\right)y_{1m}\sigma_m^p(\Lambda^{\prime}),\label{den1m-2}\\
z_{1n}\sigma_{n}^{\prime p}(\Lambda) &=&n_n^{\prime}\left[-\frac{n\phi_{\uparrow}}{\pi}+\frac{n\left(\phi_{\uparrow}-\phi_{\downarrow}\right)}{2\pi}\right]  -\int_{-\pi}^{\pi} \d k n_n^{\prime}\cos k a_{n}(\sin k-\Lambda) x_1\rho^p(k)\nonumber\\&&-\sum_{m=1}^{\infty}\int_{-\infty}^{\infty} \d \Lambda^{\prime}n_n^{\prime}\frac{1}{2\pi}\frac{\partial}{\partial \Lambda}\Theta_{nm}\left(\frac{\Lambda-\Lambda^{\prime}}{u}\right)z_{1m}\sigma_m^{\prime p}(\Lambda^{\prime})\label{den1m-3}.
\end{eqnarray}
In the following, we omit the superscript `$p$' in particle density and only use $\rho,\sigma, \sigma^{\prime}$ to denote $\rho^p,\sigma^p, \sigma^{\prime p}$, respectively. On the other hand, the derivatives of the dressed energies w.r.t. quasimomenta or rapidities can give a similar set of equations
\begin{align}
\frac{\partial \kappa(k)}{\partial k}=&2\sin k-\cos k\sum_{n=1}^{\infty} \int_{-\infty}^{\infty} \d \Lambda \left[\frac{\partial }{\partial \Lambda}a_{n}(\sin k-\Lambda)\right]\left[T\ln \left(1+\e^{-\frac{\varepsilon_n^{\prime}(\Lambda)}{T}}\right)-T\ln \left(1+\e^{-\frac{\varepsilon_n(\Lambda)}{T}}\right)\right]\nonumber\\=&2\sin k+\cos k\sum_{n=1}^{\infty} \int_{-\infty}^{\infty} \d \Lambda a_{n}(\sin k-\Lambda)\left[n_n\frac{\partial \varepsilon_n(\Lambda)}{\partial \Lambda}-n_n^{\prime}\frac{\partial \varepsilon_n^{\prime}(\Lambda)}{\partial \Lambda}\right]\label{ddren-1},\\
\frac{\partial \varepsilon_n(\Lambda)}{\partial \Lambda}=&\int_{-\pi}^{\pi}\d k \left[\frac{\partial}{\partial k} a_{n}(\sin k-\Lambda)\right] T\ln \left(1+\e^{-\frac{\kappa(k)}{T}}\right)\nonumber\\&-\sum_{m=1}^{\infty}\int_{-\infty}^{\infty} \frac{\d\Lambda^{\prime}}{2\pi}\left[\frac{\partial^2}{\partial \Lambda\partial \Lambda^{\prime}}\Theta_{nm}\left(\frac{\Lambda-\Lambda^{\prime}}{u}\right)\right]T\ln \left(1+\e^{-\frac{\varepsilon_m(\Lambda)}{T}}\right)\nonumber\\
=&\int_{-\pi}^{\pi}\d k  a_{n}(\sin k-\Lambda)n_k\frac{\partial \kappa(k)}{\partial k}-\sum_{m=1}^{\infty}\int_{-\infty}^{\infty} \frac{\d\Lambda^{\prime}}{2\pi}\left[\frac{\partial}{\partial \Lambda}\Theta_{nm}\left(\frac{\Lambda-\Lambda^{\prime}}{u}\right)\right]n_m\frac{\partial\varepsilon_m(\Lambda)}{\partial \Lambda^{\prime}}\label{ddren-2},\\
\frac{\partial\varepsilon_n^{\prime}(\Lambda)}{\partial \Lambda}=&D^{\prime}+\int_{-\pi}^{\pi}\d k  a_{n}(\sin k-\Lambda)n_k\frac{\partial \kappa(k)}{\partial k}-\sum_{m=1}^{\infty}\int_{-\infty}^{\infty} \frac{\d\Lambda^{\prime}}{2\pi}\left[\frac{\partial}{\partial \Lambda}\Theta_{nm}\left(\frac{\Lambda-\Lambda^{\prime}}{u}\right)\right]n_m^{\prime}\frac{\partial\varepsilon_m^{\prime}(\Lambda)}{\partial \Lambda^{\prime}}\label{ddren-3}.
\end{align}
By observing above two groups of equations, (\ref{den1m-1})-(\ref{den1m-3}) and (\ref{ddren-1})-(\ref{ddren-3}), we define a $(n_{\Lambda}+n_{k\Lambda}+1)$-dimensional matrix $\bm{K}$, where $n_{\Lambda}$ $(n_{k\Lambda})$ denotes the cutoff string length of considered $\Lambda$ $(k-\Lambda)$ space 
\begin{equation}
\bm{K}=
\begin{bmatrix}
0&\left[\cos ka_{n}(\sin k-\Lambda)n_n\right]|_{1\times n_{\Lambda}}&-\left[\cos ka_{n}(\sin k-\Lambda)n_m^{\prime}\right]|_{1\times n_{k\Lambda}}\\\left[a_{n}(\sin k-\Lambda)n_k\right]|_{n_{\Lambda}\times 1}&\left[-\frac{1}{2\pi}\left(\frac{\partial}{\partial \Lambda}\Theta_{nm}\left(\frac{\Lambda-\Lambda^{\prime}}{u}\right)\right)n_m\right]|_{n_{\Lambda}\times n_{\Lambda}}&\text{\Large 0}|_{n_{\Lambda}\times n_{k\Lambda}}\\\left[a_{n}(\sin k-\Lambda)n_k\right]|_{n_{k\Lambda}\times 1}&\text{\Large 0}|_{n_{k\Lambda}\times n_{\Lambda}}&\left[-\frac{1}{2\pi}\left(\frac{\partial}{\partial \Lambda}\Theta_{nm}\left(\frac{\Lambda-\Lambda^{\prime}}{u}\right)\right)n_m^{\prime}\right]|_{n_{k\Lambda}\times n_{k\Lambda}}\label{K}
\end{bmatrix}.
\end{equation}
Based on the construction of matrix $\bm{K}$, eqs.(\ref{den1m-1})-(\ref{den1m-3}) and eqs.(\ref{ddren-1})-(\ref{ddren-3}) can be cast into the form
\begin{align}
x_1\rho^p(k)=&\frac{n_k\phi_{\uparrow}}{2\pi}+\sum_{p=1}^{N}K_{p+1,1}*y_{1n}\sigma_{n}^p(\Lambda)+\sum_{p=1}^{M}K_{N+p+1,1}*z_{1n}\sigma_{n}^{\prime p}(\Lambda),\\
y_{1n}\sigma_{n}^p(\Lambda) =& -n_n\frac{n\left(\phi_{\uparrow}-\phi_{\downarrow}\right)}{2\pi}+K_{1,n+1}*x_1\rho^p(k)+\sum_{p=1}^{M}K_{p+1,n+1}*y_{1p}\sigma_{n}^p(\Lambda),\\
z_{1n}\sigma_{n}^{\prime p}(\Lambda) =&n_n^{\prime}\left[-\frac{n\phi_{\uparrow}}{\pi}+\frac{n\left(\phi_{\uparrow}-\phi_{\downarrow}\right)}{2\pi}\right]+K_{1,N+n+1}*x_1\rho^p(k)+\sum_{p=1}^{M}K_{N+p+1,N+n+1}*z_{1p}\sigma_{n}^{\prime p}(\Lambda),\\
\frac{\partial \kappa(k)}{\partial k}=&2\sin k+\sum_{p=1}^{N}K_{1,p+1}*\frac{\varepsilon_p(\Lambda)}{\partial \Lambda}+\sum_{p=1}^{M}K_{1,N+p+1}*\frac{\varepsilon_p^{\prime}(\Lambda)}{\partial \Lambda},\\
\frac{\varepsilon_n(\Lambda)}{\partial \Lambda}=&K_{n+1,1}*\frac{\partial \kappa(k)}{\partial k}+\sum_{p=1}^{M}K_{n+1,p+1}*\frac{\varepsilon_p(\Lambda)}{\partial \Lambda},\\
\frac{\partial\varepsilon_n^{\prime}(\Lambda)}{\partial \Lambda}=&D^{\prime}+K_{N+n+1,1}*\frac{\partial \kappa(k)}{\partial k}+\sum_{p=1}^{M}K_{N+n+1,N+p+1}*\frac{\varepsilon_p^{\prime}(\Lambda)}{\partial \Lambda}.
\end{align}
The above equations manifest highly symmetrical forms and we can generalise them to normal style
\begin{eqnarray}
f_1&=&f_1^{(0)}+\sum_{p=1}^{N}K_{1,p+1}*f_{2p}+\sum_{p=1}^{M}K_{1,N+p+1}*f_{3p},\\f_{2n}&=&f_{2n}^{(0)}+K_{n+1,1}*f_1+\sum_{p=1}^{M}K_{n+1,p+1}*f_{2p},\\f_{3n}&=&f_{3n}^{(0)}+K_{N+n+1,1}*f_1+\sum_{p=1}^{M}K_{N+n+1,N+p+1}*f_{3p},\\g_1&=&g_1^{(0)}+\sum_{p=1}^{N}K_{p+1,1}*g_{2p}+\sum_{p=1}^{M}K_{N+p+1,1}*g_{3p},\\g_{2n}&=&g_{2n}^{(0)}+K_{1,n+1}*g_1+\sum_{p=1}^{M}K_{p+1,n+1}*g_{2p},\\g_{3n}&=&g_{3n}^{(0)}+K_{1,N+n+1}*g_1+\sum_{p=1}^{M}K_{N+p+1,N+n+1}*g_{3p}.
\end{eqnarray}
Based on above general expressions, it can be proven
\begin{equation}
\int f_1^{(0)}g_1+\sum_n \int f_{2n}^{(0)}g_{2n}+\sum_n \int f_{3n}^{(0)}g_{3n}=\int f_1g_1^{(0)}+\sum_n \int f_{2n}g_{2n}^{(0)}+\sum_n \int f_{3n}g_{3n}^{(0)}.
\end{equation}
Thus $E_1$ in eq.(\ref{E1}) can be readily shown to be zero as
\begin{eqnarray}
E_1&=&\int_{-\pi}^{\pi}\d k\frac{n_k\phi_{\uparrow}}{2\pi}\frac{\partial \kappa(k)}{\partial k}+\sum_{n=1}^{\infty} \int_{-\infty}^{\infty} \d \Lambda(-n_n)\frac{n\left(\phi_{\uparrow}-\phi_{\downarrow}\right)}{2\pi}\frac{\partial \varepsilon_n(\Lambda)}{\partial \Lambda}\no\\&&
+\sum_{n=1}^{\infty} \int_{-\infty}^{\infty} \d \Lambda n_n^{\prime}\left[-\frac{n\phi_{\uparrow}}{\pi}+\frac{n\left(\phi_{\uparrow}-\phi_{\downarrow}\right)}{2\pi}\right]\frac{\partial\varepsilon_n^{\prime}(\Lambda)}{\partial \Lambda}=0.
\end{eqnarray}

\subsubsection{{Evaluation of $E_2$}}
In order to simplify the second-order energy $E_2$ in eq.(\ref{E2}), we first rewrite eqs.(\ref{den2}) as 
\begin{eqnarray}
&&\rho x_2-\frac{1}{2}\frac{\partial}{\partial k}(\rho x_1^2)\no\\&=&\sum_n\int \d \Lambda\frac{a_n(\sin k-\Lambda)}{1+\exp(\kappa/T)}\left[\left(y_{2n}\sigma_n-\frac{1}{2}\frac{\partial}{\partial \Lambda}(y_{1n}^2\sigma_n)\right)+\left(z_{2n}\sigma_n^{\prime}-\frac{1}{2}\frac{\partial}{\partial \Lambda}(z_{1n}^2\sigma_n^{\prime})\right)\right]\no\\&&+\frac{\rho x_1^2}{1+\exp(-\kappa/T)}\frac{1}{2T}\frac{\partial \kappa}{\partial k},\\
&&\sigma_ny_{2n}-\frac{1}{2}\frac{\partial}{\partial \Lambda}(\sigma_ny_{1n}^2)\no\\&=&-\sum_m\int \d\Lambda^{\prime} \frac{1}{1+\exp(\varepsilon_n/T)}\frac{1}{2\pi}\frac{\partial}{\partial \Lambda}\Theta_{nm}\left(\frac{\Lambda-\Lambda^{\prime}}{u}\right)\left(y_{2m}\sigma_m-\frac{1}{2}\frac{\partial}{\partial \Lambda^{\prime}}(y_{1m}^2\sigma_m)\right)\nonumber\\&&+\frac{\sigma_ny_{1n}^2}{1+\exp(-\varepsilon_n/T)}\frac{1}{2T}\frac{\partial \varepsilon_n}{\partial \Lambda}+\int \d k\frac{a_n(\sin k-\Lambda)\cos k}{1+\exp(\varepsilon_n/T)}\left[\rho x_2-\frac{1}{2}\frac{\partial}{\partial k}(\rho x_1^2)\right],\\
&&\sigma_n^{\prime}z_{2n}-\frac{1}{2}\frac{\partial}{\partial \Lambda}(\sigma_n^{\prime}z_{1n}^2)\no\\&=&-\sum_m\int \d\Lambda^{\prime} \frac{1}{1+\exp(\varepsilon_n^{\prime}/T)}\frac{1}{2\pi}\frac{\partial}{\partial \Lambda}\Theta_{nm}\left(\frac{\Lambda-\Lambda^{\prime}}{u}\right)\left(z_{2m}\sigma_m^{\prime}-\frac{1}{2}\frac{\partial}{\partial \Lambda^{\prime}}(z_{1m}^2\sigma_m^{\prime})\right)\nonumber\\&&+\frac{\sigma_n^{\prime}z_{1n}^2}{1+\exp(-\varepsilon_n^{\prime}/T)}\frac{1}{2T}\frac{\partial \varepsilon_n^{\prime}}{\partial \Lambda}-\int \d k\frac{a_n(\sin k-\Lambda)\cos k}{1+\exp(\varepsilon_n^{\prime}/T)}\left[\rho x_2-\frac{1}{2}\frac{\partial}{\partial k}(\rho x_1^2)\right].
\end{eqnarray}
Therefore, $E_2$ can be expressed as
\begin{eqnarray}
E_2&=&\int_{-\pi}^{\pi}\d k 2\sin k\left[\rho x_2-\frac{1}{2}\frac{\partial}{\partial k}(\rho x_1^2)\right]+\sum_{n=1}^{\infty}\int_{-\infty}^{\infty} \d \Lambda D^{\prime} \left[z_{2n}\sigma_n^{\prime}-\frac{1}{2}\frac{\partial}{\partial \Lambda}(z_{1n}^2\sigma_n^{\prime})\right]\nonumber\\&=&\int_{-\pi}^{\pi}\d k \frac{\rho x_1^2}{1+\exp(-\kappa/T)}\frac{1}{2T}\left(\frac{\partial \kappa}{\partial k}\right)^2+\sum_{n=1}^{\infty}\int_{-\infty}^{\infty} \d \Lambda \frac{\sigma_ny_{1n}^2}{1+\exp(-\varepsilon_n/T)}\frac{1}{2T}\left(\frac{\partial \varepsilon_n}{\partial \Lambda}\right)^2\nonumber\\&&+\sum_{n=1}^{\infty}\int_{-\infty}^{\infty} \d \Lambda\frac{\sigma_n^{\prime}z_{1n}^2}{1+\exp(-\varepsilon_n^{\prime}/T)}\frac{1}{2T}\left(\frac{\partial \varepsilon_n^{\prime}}{\partial \Lambda}\right)^2,
\end{eqnarray}
where the first line arises from integration by parts. Having obtained explicit expressions of $E_0,E_1,E_2$, it's easy to obtain the expressions of DWs using eqs.(\ref{Dc0}) and (\ref{Ds0}). Note that only second-order energy $E_2$ depends on flux $\phi$, thus DWs take universal simple forms
\begin{eqnarray}
D^{\alpha \beta}&=&\frac{1}{2}\left\langle\frac{\partial^2E_2}{\partial \phi^{\alpha}\phi^{\beta}}\right\rangle\nonumber\\&=&\frac{1}{2T}\int_{-\pi}^{\pi}\d k \frac{1}{1+\exp(-\kappa/T)}\left(\frac{\partial \kappa}{\partial k}\right)^2\rho\frac{\d x_1}{\d\phi^{\alpha}}\frac{\d x_1}{\d\phi^{\beta}}\no\\&&+\frac{1}{2T}\sum_{n=1}^{\infty}\int_{-\infty}^{\infty} \d \Lambda \frac{1}{1+\exp(-\varepsilon_n/T)}\left(\frac{\partial \varepsilon_n}{\partial \Lambda}\right)^2\sigma_n\frac{\d y_{1n}}{\d\phi^{\alpha}}\frac{\d y_{1n}}{\d\phi^{\beta}}\nonumber\\&&+\frac{1}{2T}\sum_{n=1}^{\infty}\int_{-\infty}^{\infty} \d \Lambda\frac{1}{1+\exp(-\varepsilon_n^{\prime}/T)}\left(\frac{\partial \varepsilon_n^{\prime}}{\partial \Lambda}\right)^2\sigma_n^{\prime}\frac{\d z_{1n}}{\d\phi^{\alpha}}\frac{\d z_{1n}}{\d\phi^{\beta}}\label{D-cs0}.
\end{eqnarray}
The difference of charge and spin DWs is reflected in the derivatives of finite-size rapidities w.r.t. fluxes, namely the terms $dx_1(y_{1n},z_{1n})/d\phi$. In terms of eqs.(\ref{den1}), $dx_1(y_{1n},z_{1n})/d\phi$ are given by
\begin{eqnarray}
2\pi(\rho+\rho^h)\frac{dx_1}{d\phi}&=&g_k+2\pi\sum_{n=1}^{\infty}\int_{-\infty}^{\infty} \d \Lambda a_n(\sin k-\Lambda)\left(\sigma_n\frac{dy_{1n}}{d\phi}+\sigma_n^{\prime}\frac{dz_{1n}}{d\phi}\right)\label{q-dr},\\
2\pi(\sigma_n+\sigma_n^h)\frac{dy_{1n}}{d\phi}&=&g_{\Lambda}+2\pi\int_{-\pi}^{\pi}\d ka_n(\Lambda-\sin k)\cos k \rho\frac{dx_1}{d\phi}\no\\&&-\sum_{m=1}^{\infty}\int \d\Lambda^{\prime}\frac{\partial}{\partial \Lambda}\Theta_{nm}\left(\frac{\Lambda-\Lambda^{\prime}}{u}\right)\sigma_m^{\prime}\frac{dy_{1m}}{d\phi},\\
2\pi(\sigma_n^{\prime}+\sigma_n^{\prime h})\frac{dz_{1n}}{d\phi}&=&g_{k\Lambda}-2\pi\int_{-\pi}^{\pi}\d ka_n(\Lambda-\sin k)\cos k \rho\frac{dx_1}{d\phi}\no\\&&-\sum_{m=1}^{\infty}\int \d\Lambda^{\prime}\frac{\partial}{\partial \Lambda}\Theta_{nm}\left(\frac{\Lambda-\Lambda^{\prime}}{u}\right)\sigma_m^{\prime}\frac{dz_{1m}}{d\phi},
\end{eqnarray}
in which $g_k,g_{\Lambda},g_{k\Lambda}$ are constants, 
\begin{equation}
\begin{aligned}
\text{for charge } &:g_k=1,g_{\Lambda}=0,g_{k\Lambda}=-2n_{\Lambda^{'}_n},\\ 
\text{for spin } &:g_k=1,g_{\Lambda}=-2n_{\Lambda_n},g_{k\Lambda}=0\label{f-l},
\end{aligned}
\end{equation}
respectively. Define the velocities of various quasiparticles, 
\begin{eqnarray}
v_k&=&\frac{1}{2\pi(\rho+\rho^h)}\frac{\partial \kappa}{\partial k},\\
v_{\Lambda}&=&\frac{1}{2\pi(\sigma_n+\sigma_n^h)}\frac{\partial \varepsilon_n}{\partial \Lambda},\\
v_{k\Lambda}&=&\frac{1}{2\pi(\sigma_n^{\prime}+\sigma_n^{\prime h})}\frac{\partial \varepsilon_n^{\prime}}{\partial \Lambda},
\end{eqnarray}
thus the expressions of DWs of eq.(\ref{D-cs0}) can be rewritten as 
\begin{eqnarray}
D^{\alpha \beta}&=&\frac{1}{2T}\sum_{\gamma=k,\Lambda,k-\Lambda}\int d\theta_{\gamma}\rho_{\gamma}(1-n_{\gamma})v_{\gamma}^2\left(2\pi(\rho_{\gamma}+\rho_{\gamma}^h)\frac{dx_{\gamma,1}}{d\phi^{\alpha}}\right)\left(2\pi(\rho_{\gamma}+\rho_{\gamma}^h)\frac{dx_{\gamma,1}}{d\phi^{\beta}}\right)\label{D-csm},
\end{eqnarray}
in which $\rho_{\gamma},\rho_{\gamma}^h$ denote the particle density and hole density of $\gamma$-type particle, $n_{\gamma}=\rho_{\gamma}/(\rho_{\gamma}+\rho_{\gamma}^h)$ is the occupation number. The eq.(\ref{D-csm}) is derived in terms of finite size correction under the twisted boundary condition. On the other hand, hydrodynamic method can also give the results of DW based on generalized Gibbs ensemble
\begin{eqnarray}
D^{\alpha \beta}=\int d\theta \rho(\theta)(1-n(\theta))v^{\eff}(\theta)^2{q^{\alpha}}^{\dr}(\theta){q^{\beta}}^{\dr}(\theta)\label{D-hy},
\end{eqnarray}
in which $q^{\dr}$ is the dressed charge after conducting dressing transformation on local conserved charge $q$ and $v^{\eff}$ is effective velocity which is defined as
\begin{equation}
v^{\eff}=\frac{(e^{\prime})^{\dr}(\theta)}{|(p^{\prime})^{\dr}(\theta)|}=\frac{(e^{\ph})^{\prime}}{|(p^{\ph})^{\prime}|},
\end{equation}
where $e,p$ are bare energy and bare momentum, $e^{\ph},p^{\ph}$ are physical energy and momentum. Comparing eq.(\ref{D-csm}) with eq.(\ref{D-hy}), we can identify the following relation
\begin{equation}
{q^{\alpha}}^{\dr}=\pm2\pi(\rho_{\gamma}+\rho_{\gamma}^h)\frac{dx_{\gamma,1}}{d\phi^{\alpha}}\label{pm},
\end{equation}
with the plus and minus sign to be determined, see eq.(\ref{sign}) below.  

\subsection{Thermodynamic Bethe ansatz equations and Dressing transformation}
In the presence of external potentials, the Hamiltonian of 1D Hubbard model is give by 
\begin{equation}
H=-\sum_{j=1}^{L} \sum_{a=\uparrow \downarrow}\left(c_{j, a}^{+} c_{j+1, a}+c_{j+1, a}^{+} c_{j, a}\right)+u \sum_{j=1}^{L} (1-2n_{j \uparrow}) (1-2n_{j \downarrow})-\mu\hat{N}-2B\hat{S}_z. \label{eq-H}
\end{equation}
In terms of Yang-Yang thermodynamic method, the relevant TBA equations can be shown as
\begin{equation}
\ln Y_{a}=\beta e_a- \bm{A }_{ab}*\ln(1+1/{Y_b})\label{Y},
\end{equation}
in which repeated indice $b$ for three types of quasiparticles $k,\Lambda,k-\Lambda$ represents Einstein summation. We define a $(n_{\Lambda}+n_{k\Lambda}+1)$-dimensional matrix $\bm{A}$, where $n_{\Lambda}(n_{k\Lambda})$ have the same meaning as eq.(\ref{K})
\begin{equation}
\bm{A}=
\begin{bmatrix}
0&\left[a_{n}(\sin k-\Lambda)\right]|_{1\times n_{\Lambda}}&-\left[a_{n}(\sin k-\Lambda)\right]|_{1\times n_{k\Lambda}}\\\left[\cos ka_{n}(\sin k-\Lambda)\right]|_{n_{\Lambda}\times 1}&\left[-\frac{1}{2\pi}\left(\frac{\partial}{\partial \Lambda}\Theta_{nm}\left(\frac{\Lambda-\Lambda^{\prime}}{u}\right)\right)\right]|_{n_{\Lambda}\times n_{\Lambda}}&\text{\Large 0}|_{n_{\Lambda}\times n_{k\Lambda}}\\\left[\cos ka_{n}(\sin k-\Lambda)\right]|_{n_{k\Lambda}\times 1}&\text{\Large 0}|_{n_{k\Lambda}\times n_{\Lambda}}&\left[-\frac{1}{2\pi}\left(\frac{\partial}{\partial \Lambda}\Theta_{nm}\left(\frac{\Lambda-\Lambda^{\prime}}{u}\right)\right)\right]|_{n_{k\Lambda}\times n_{k\Lambda}},\label{A}
\end{bmatrix}.
\end{equation}
And $e_a$ denotes quasiparticles' bare energies
\begin{eqnarray}
e_k&=&-2\cos k-\mu-2u-B,\\
e_{n|\Lambda}&=&2nB,\\
e_{n|k\Lambda}&=&4 \textmd{Re} \sqrt{1-(\Lambda-\mathrm{i} n u)^{2}}-2 n \mu-4 n u.
\end{eqnarray}
The physical energy $e^{\ph}$ of each particle is 
\begin{equation}
e^{\ph}_a(\theta)=\ln Y_{a}/{\beta}.
\end{equation}
$Y_a$ in eq.(\ref{Y}) is related to density via $Y_a=\rho_a^h/{\rho_a}$. Thus the occupation number is
\begin{equation}
n_a=\frac{\rho_a}{\rho_a+\rho_a^h}=\frac{1}{1+Y_a}.
\end{equation}
Meanwhile, the root densities of particles and holes are given by
\begin{equation}
\rho_a+\rho_a^h=\tau_ap^{\prime}(\theta)/(2\pi)+ \bm{A^T}_{ab}*\rho_b\label{rho},
\end{equation}
where $\tau_a=\mathrm{sign}(p_a^{\prime}(\theta))=1,1,-1$ for $k,\Lambda,k-\Lambda$ respectively, and $p^{\prime}=k^{\prime}$ denotes the derivative of quasimomentum w.r.t. rapidity $\theta$ 
\begin{eqnarray}
p^{\prime}_k&=&1,\\
p^{\prime}_{n|\Lambda}&=&0,\\
p^{\prime}_{n|k\Lambda}&=&-2 \textmd{Re}\left[1/\sqrt{1-(\Lambda-\mathrm{i} n u)^{2}}\right]. 
\end{eqnarray}
Correspondingly, the physical momentum $p^{\ph}$ of each particle can be related to densities via
\begin{equation}
p^{\ph}_a(\theta)=\tau_a2\pi\int_0^{\theta}d\theta^{\prime}(\rho_a+\rho_a^h)+\delta_{a,n|k\Lambda}\pi(n+1).
\end{equation}
It can be shown from eq.(\ref{Y}) and (\ref{rho}) that in interacting system, the energy and momentum of single quasiparticle should be dressed by interaction. The dressing transformation is defined as
\begin{equation}
f_a^{\dr}=\left(\textmd{I}-\bm{B}\right)_{ab}^{-1}*f_b\label{dr},
\end{equation}
for arbitrary bare quantity $f_b$ with matrix $\bm{B}$
\begin{equation}
\bm{B}=\!\!
\begin{bmatrix}
0\!\!&\!\!\left[a_{n}(\sin k-\Lambda)n_n\right]|_{1\times n_{\Lambda}}\!\!&\!\!-\left[a_{n}(\sin k-\Lambda)n_m^{\prime}\right]|_{1\times n_{k\Lambda}}\\\cos k\left[a_{n}(\sin k-\Lambda)n_k\right]|_{n_{\Lambda}\times 1}\!\!\!\!&\!\!\!\!\left[-\frac{1}{2\pi}\left(\frac{\partial}{\partial \Lambda}\Theta_{nm}\left(\frac{\Lambda-\Lambda^{\prime}}{u}\right)\right)n_m\right]|_{n_{\Lambda}\times n_{\Lambda}}\!\!\!\!&\!\!\!\!\text{\Large 0}|_{n_{\Lambda}\times n_{k\Lambda}}\\\cos k\left[a_{n}(\sin k-\Lambda)n_k\right]|_{n_{k\Lambda}\times 1}\!\!&\!\!\text{\Large 0}|_{n_{k\Lambda}\times n_{\Lambda}}\!\!\!\!&\!\!\!\!\left[-\frac{1}{2\pi}\left(\frac{\partial}{\partial \Lambda}\Theta_{nm}\left(\frac{\Lambda-\Lambda^{\prime}}{u}\right)\right)n_m^{\prime}\right]|_{n_{k\Lambda}\times n_{k\Lambda}}\label{B}
\end{bmatrix}.
\end{equation}
In fact, $\bm{B}$ and $\bm{A}$ satisfy $\bm{B}_{ab}=\bm{A}_{ab}n_b$. Thus, we can bulid the relations between physical quantities and bare quantities
\begin{equation}
(e^{\prime})^{\dr}=(e^{\ph})^{\prime},(p^{\prime})^{\dr}=(p^{\ph})^{\prime}.
\end{equation}
	
Considering the Hamiltonian displayed in eq.(\ref{eq-H}), the system conserves particle number $o$ and magnetization $m$, associated with the two U(1) symmetries. In terms of the patterns of Bethe roots, we can identify each bare quantity corresponding to these three configurations, listed below
\begin{equation}
\begin{aligned}
k&:&o_k&=1&m_k&=1/2,\\
\Lambda &:& o_{n|\Lambda}&=0&m_{n|\Lambda}&=-n_{\Lambda_n},\\
k-\Lambda &:& o_{n|k-\Lambda}&=2n_{\Lambda^{'}_n}&m_{n|k-\Lambda}&=0\label{hy-l}.
\end{aligned} 
\end{equation}
The above bare quantities correspond well to eq.(\ref{f-l}) except the missing minus sign in $k-\Lambda$ string. Considering local conserved particles emerged in integrable system, based on eqs.(\ref{Y}) and (\ref{dr}), it's obvious to relate bare charges to dressed charges
\begin{eqnarray}
o_a^{\dr}&=&\frac{\partial \ln Y_a}{\partial(-\beta \mu)}\label{qdr},\\
m_a^{\dr}&=&\frac{\partial \ln Y_a}{\partial(-2\beta B)}\label{mdr},\\
e_a^{\dr}&=&\frac{\partial \ln Y_a}{\partial\beta}.
\end{eqnarray}
Comparing eq.(\ref{q-dr}) with eqs.(\ref{qdr}) and (\ref{mdr}), we can deduce the correct sign in eq.(\ref{pm})
\begin{equation}
q_a^{\dr}=\mathrm{sign}(p_a^{\prime})2\pi(\rho_{a}+\rho_{a}^h)\frac{dx_{a}}{d\phi}. \label{sign}
\end{equation}	
Similarly, for attractive Hubbard model, $\Lambda$ string needs minus sign due to $\mathrm{sign}(p_{n|\Lambda}^{\prime})=-1$.

\subsection{Discussions of dressed charges at $T=0,B=0$}
In the limit of zero temperature, only charge and unbound spin degrees can survive. Thus, the dressed charges can be simplified as
\begin{eqnarray}
q_k^{\dr}&=&1+\int_{-A}^{A} \d \Lambda a_1(\sin k-\Lambda)q_{\Lambda}^{\dr}\label{qc-dr},\\
q_{\Lambda}^{\dr}&=&g+\int_{-Q}^{Q}\d k \cos k a_1(\Lambda-\sin k) q_k^{\dr}-\int_{-A}^{A} \d\Lambda^{\prime}a_2(\Lambda-\Lambda^{\prime})q_{\Lambda}^{\dr}\label{qs-dr},
\end{eqnarray}
where $g=0,-2$ for charge and spin transport. Subscripts with $k,\Lambda$ mean the contributions from $k$ and length-1 $\Lambda$ string. At vanishing magnetic field, by conducting the Fourier transformation of rapidity $\Lambda$, the spin degree has the form
\begin{equation}
	q_{\Lambda}^{\dr}(w)=\frac{2\pi g\delta(w)}{1+\e^{-2u|w|}}+\int_{-Q}^{Q}\d k \cos k q_k^{\dr}(k)\frac{\e^{\mi w\sin k}}{2\cosh(uw)}.
\end{equation}
Then taking the inverse Fourier transformation, the spin part is expressed as 
\begin{equation}
	q_{\Lambda}^{\dr}(\Lambda)=\frac{g}{2}+\int_{-Q}^{Q}\d k \frac{\cos k q_k^{\dr}(k)}{4u\cosh(\pi(\Lambda-\sin k)/2u)}\label{q-l}.
\end{equation}
In terms of above expressions, the final result of charge part is given by
\begin{equation}
	q_k^{\dr}(k)=1+\frac{g}{2}+\int_{-Q}^{Q}\d k^{\prime}\cos k^{\prime} q_k^{\dr}(k^{\prime})R(\sin k^{\prime}-\sin k)\label{q-k},
\end{equation}	
where the function $R(x)$ is defined as
\begin{equation}
	R(x)=\frac{1}{2\pi}\int_{-\infty}^{\infty}\d w \frac{\e^{\mi wx}}{1+\e^{2u|w|}}.
\end{equation}

\section{Linear U(1) Drude weights and Tomonaga-Luttinger Liquid (TLL)}

\subsection{General properties at TLL region}
Using eq.(\ref{D-csm}), we can analyze the universal properties of the DWs. Firstly, we show the exact results in TLL phases. In the low temperature regime, only real $k$ and length-1 $\Lambda$ string can make contributions. Due to the same expression of each type particle in the summation of DWs $D^{\alpha\beta}=D^{\alpha\beta}_{k}+ D^{\alpha\beta}_{\Lambda}$, we only give the derivation of $k$  quasiparticle ($D^{\alpha\beta}_{k}$). Taking charge DW as an example, the expression $D^{cc}_{k}$ can be dealt with
\begin{eqnarray}
	D^{cc}_{k}&=&\frac{1}{2T}\int_{-\pi}^{\pi} dk\rho_k(1-n_{k})v_{k}^2\left(2\pi(\rho_{k}+\rho_{k}^h)\frac{dx_{1}}{d\phi^c}\right)^2\nonumber\\
	&=&\int_{-\pi}^{\pi} \frac{\d k}{4\pi}\frac{\d}{\d k}\left(\frac{-1}{1+\e^{\kappa/T}}\right){{q_k^c}^{\dr}}^2v_k\nonumber\\
	&=&\int_{-\pi}^{\pi} \frac{1}{4\pi}\frac{1}{1+\e^{\kappa/T}}\d ({{q_k^c}^{\dr}}^2v_k)\nonumber\\
	&=&\int_{\kappa(0)}^{\kappa(\pi)} \frac{1}{2\pi}\frac{1}{1+\e^{\kappa/T}}\frac{\d}{\d \kappa}({{q_k^c}^{\dr}}^2v_k)\d \kappa\nonumber\\
	&=&\int_{\kappa(0)/T}^{\kappa(\pi)/T} \frac{1}{2\pi}\frac{1}{1+\e^x}\frac{\d}{\d x}[{{q_k^c}^{\dr}}^2(Tx)v_k(Tx)]\d x\nonumber\\
	&=&\left[\int_{\kappa(0)/T}^{\kappa(Q)/T}+\int_{\kappa(Q)/T}^{\kappa(\pi)/T}\right] \frac{1}{2\pi}\frac{1}{1+\e^x}\frac{\d}{\d x}[{{q_k^c}^{\dr}}^2(Tx)v_k(Tx)]\d x\nonumber\\
	&=&\left[\int_{-\infty}^{0}\left(1-\frac{1}{1+\e^{-x}}\right)+\int_{0}^{\infty}\frac{1}{1+\e^x}\right] \frac{1}{2\pi}\frac{\d}{\d x}[{{q_k^c}^{\dr}}^2(Tx)v_k(Tx)]\d x\nonumber\\
	&=&\frac{1}{2\pi}{{q_k^c}^{\dr}}^2(Q)v_k(Q)\nonumber\\&&+\int_{0}^{\infty}\frac{1}{2\pi}\frac{1}{1+\e^x}\left\{\frac{\d}{\d (Tx)}[{{q_k^c}^{\dr}}^2(Tx)v_k(Tx)]-\frac{\d}{\d (-Tx)}[{{q_k^c}^{\dr}}^2(-Tx)v_k(-Tx)]\right\}T\d x\nonumber\\
	&=&\frac{1}{2\pi}{{q_k^c}^{\dr}}^2(Q)v_k(Q)+\int_{0}^{\infty}\frac{1}{2\pi}\frac{2Tx}{1+\e^x}\frac{\d^2}{\d \kappa^2}({{q_k^c}^{\dr}}^2(\kappa)v_k(\kappa))|_{\kappa=0}T\d x\nonumber\\
	&=&\frac{1}{2\pi}{{q_k^c}^{\dr}}^2(Q)v_k(Q)+\frac{\pi T^2}{12}\frac{\d^2}{\d \kappa^2}({{q_k^c}^{\dr}}^2(\kappa)v_k(\kappa))\Big|_{\kappa=0}\label{TLL},
\end{eqnarray}
where $Q$ denotes the Fermi point of the charge sector. Similarly, the charge contribution to the crossed DW can be expressed as 
\begin{equation}
D^{cs}_{k}=\frac{1}{2\pi}{{q_k^{c}}^{\dr}}(Q){{q_k^{s}}^{\dr}}(Q)v_k(Q)+\frac{\pi T^2}{12}\frac{\d^2}{\d \kappa^2}({{q_k^{c}}^{\dr}}(\kappa){{q_k^{s}}^{\dr}}(\kappa)v_k(\kappa))\Big|_{\kappa=0}\label{TLLcc1}.
\end{equation}

\subsection{General properties in quantum criticality (QC) region}
In this part, we give universal scaling laws of DWs at arbitrary potentials and interactions in the QC region.
In fact, the universal laws around the quantum critical points are related to the densities and dressed charges of the ground state. The five phases, labelled I to V, are shown in Fig.~1 of the main text. We first present the general results
\begin{eqnarray}
	\textbf{I-II: }D_k^{\alpha\beta}&=&-\frac{1}{\rho(0)}\left(\frac{{q_k^{\alpha}}^{\dr}(0)}{2\pi}\right)\left(\frac{{q_k^{\beta}}^{\dr}(0)}{2\pi}\right)\left(\frac{\kappa^{\prime \prime}(0)}{2}\right)^{\frac{1}{2}}\pi^{\frac{1}{2}}T^{\frac{1}{2}}\Li_{\frac{1}{2}}\left(-\e^{\frac{-\kappa(0)}{T}}\right)\label{12},\\
	\textbf{II-III: }D_k^{\alpha\beta}&=&-\frac{1}{\rho(\pi)}\left(\frac{{q_k^{\alpha}}^{\dr}(\pi)}{2\pi}\right)\left(\frac{{q_k^{\beta}}^{\dr}(\pi)}{2\pi}\right)\left(\frac{-\kappa^{\prime \prime}(\pi)}{2}\right)^{\frac{1}{2}}\pi^{\frac{1}{2}}T^{\frac{1}{2}}\Li_{\frac{1}{2}}\left(-\e^{\frac{\kappa(\pi)}{T}}\right)\label{23},\\
	\textbf{III-V: }D_{\Lambda}^{\alpha\beta}&=&-\frac{1}{\sigma(0)}\left(\frac{{q_{\Lambda}^{\alpha}}^{\dr}(0)}{2\pi}\right)\left(\frac{{q_{\Lambda}^{\beta}}^{\dr}(0)}{2\pi}\right)\left(\frac{\varepsilon_1^{\prime \prime}(0)}{2}\right)^{\frac{1}{2}}\pi^{\frac{1}{2}}T^{\frac{1}{2}}\Li_{\frac{1}{2}}\left(-\e^{\frac{-\varepsilon_1(0)}{T}}\right)\label{35},\\
	\textbf{II-IV: }D_{\Lambda}^{\alpha\beta}&=&-\frac{1}{\sigma(0)}\left(\frac{{q_{\Lambda}^{\alpha}}^{\dr}(0)}{2\pi}\right)\left(\frac{{q_{\Lambda}^{\beta}}^{\dr}(0)}{2\pi}\right)\left(\frac{\varepsilon_1^{\prime \prime}(0)}{2}\right)^{\frac{1}{2}}\pi^{\frac{1}{2}}T^{\frac{1}{2}}\Li_{\frac{1}{2}}\left(-\e^{\frac{-\varepsilon_1(0)}{T}}\right)\label{24s},\\
	D_{k}^{\alpha\beta}&=&D^0\left[1-f_{\frac{1}{2}}\sum_{e=\alpha, \beta}\frac{\frac{1}{2}\sigma(0){{q_{\Lambda}^{e}}^{\dr}}(0)^{\prime}-\sigma(0)^{\prime}{q_{\Lambda}^{e}}^{\dr}(0)}{\rho^0{{q_{k}^{e}}^{\dr}}^0}\right]+\frac{\pi^2T^2}{6}\frac{\partial^2}{\partial \kappa^2}D^0\Big|_{\kappa(k)=0}\label{24c},\\
	\textbf{IV-V: }D_k^{\alpha\beta}&=&-\frac{1}{\rho(\pi)}\left(\frac{{q_k^{\alpha}}^{\dr}(\pi)}{2\pi}\right)\left(\frac{{q_k^{\beta}}^{\dr}(\pi)}{2\pi}\right)\left(\frac{-\kappa^{\prime \prime}(\pi)}{2}\right)^{\frac{1}{2}}\pi^{\frac{1}{2}}T^{\frac{1}{2}}\Li_{\frac{1}{2}}\left(-\e^{\frac{\kappa(\pi)}{T}}\right)\label{45c},\\
	D_{\Lambda}^{\alpha\beta}&=&D^0\left[1+k_{\frac{1}{2}}\sum_{e=\alpha, \beta}\frac{\frac{1}{2}\rho(\pi){{q_{k}^{e}}^{\dr}}(\pi)^{\prime}-\rho(\pi)^{\prime}{q_{k}^{e}}^{\dr}(\pi)}{\sigma^0{{q_{\Lambda}^{e}}^{\dr}}^0}\right]+\frac{\pi^2T^2}{6}\frac{\partial^2}{\partial \varepsilon_1^2}D^0\Big|_{\varepsilon_1(\Lambda)=0}\label{45s},
\end{eqnarray}
where $\rho(0)(q_k^{\dr}(0)),\rho(\pi)(q_k^{\dr}(\pi)),\sigma(0)(q_{\Lambda}^{\dr}(0))$ denote charge density (charge component of dressed charge) at $k=0,k=\pi$ and spin density (spin component of dressed spin) at $\Lambda=0$. $\rho^0,\sigma^0,{q_{\Lambda}^{\dr}}^0,{q_{k}^{\dr}}^0$ denote the values at ground state.  ${{q_{\Lambda}^{e}}^{\dr}}(0)^{\prime},\sigma(0)^{\prime} ({{q_{k}^{e}}^{\dr}}(\pi)^{\prime},\rho(\pi)^{\prime})$ mean the derivatives w.r.t. charge (spin) Fermi point, and $\prime\prime$ means the second derivative w.r.t. relevant rapidities. $D^0$ denotes background contribution at zero temperature, $f_{\frac{1}{2}},k_{\frac{1}{2}}$ are given by
\begin{eqnarray}
	f_{\frac{1}{2}}&=&-\frac{1}{2}\left(\frac{\varepsilon_1^{\prime \prime}(0)}{2}\right)^{-\frac{1}{2}}\pi^{\frac{1}{2}}T^{\frac{1}{2}}\Li_{\frac{1}{2}}\left(-\e^{\frac{-\varepsilon_1(0)}{T}}\right),\\
	k_{\frac{1}{2}}&=&-\frac{1}{2}\left(\frac{-\kappa^{\prime \prime}(\pi)}{2}\right)^{-\frac{1}{2}}\pi^{\frac{1}{2}}T^{\frac{1}{2}}\Li_{\frac{1}{2}}\left(-\e^{\frac{\kappa(\pi)}{T}}\right).
\end{eqnarray}

Below we give the proofs for II-IV and IV-V phase transitions, i.e., eqs.(\ref{24s})$\sim$(\ref{45s}), since these two transitions involve both charge and spin degrees of freedom, whereas other transitions involve only charge (I-II and II-III) or spin (III-V) degrees of freedom and hence are relatively simpler. First we denote $\widetilde{g}=g/(2\pi)$,
$\widetilde{q^{\dr}}=q^{\dr}/(2\pi)$ in the following calculation. And we present 
the equations of density $\rho^t (=\rho+\rho^h),\sigma^t (=\sigma+\sigma^h)$ and dressed charge $\widetilde{q_k^{\dr}},\widetilde{q_{\Lambda}^{\dr}}$ here
 \begin{eqnarray}\left\{\begin{aligned}
	\rho^t&=\frac{1}{2\pi}+\cos k\int_{-\infty}^{\infty}\d \Lambda a_1(\sin k-\Lambda) \frac{\sigma^t}{1+\e^{\varepsilon_1/T}},\\\sigma^t&=-\int_{-\infty}^{\infty}\d \Lambda^{\prime} a_2(\Lambda-\Lambda^{\prime}) \frac{\sigma^t}{1+\e^{\varepsilon_1/T}}+\int_{-\pi}^{\pi}\d k a_1(\sin k-\Lambda) \frac{\rho^t}{1+\e^{\kappa/T}}\label{dt},
	\end{aligned}\right.\\
	\left\{\begin{aligned}
	\widetilde{q_k^{\dr}}&=\frac{1}{2\pi}+\int_{-\infty}^{\infty}\d \Lambda a_1(\sin k-\Lambda) \frac{\widetilde{q_{\Lambda}^{\dr}}}{1+\e^{\varepsilon_1/T}},\\
	\widetilde{q_{\Lambda}^{\dr}}&=	\widetilde{g}-\int_{-\infty}^{\infty}\d \Lambda^{\prime} a_2(\Lambda-\Lambda^{\prime}) \frac{\widetilde{q_{\Lambda}^{\dr}}}{1+\e^{\varepsilon_1/T}}+\int_{-\pi}^{\pi}\d k a_1(\sin k-\Lambda)\cos k \frac{\widetilde{q_k^{\dr}}}{1+\e^{\kappa/T}}\label{dct}.
	\end{aligned}\right.
\end{eqnarray}
Note that $\kappa(k)=-2\cos k+2C_1\sin^2k+C_2,\varepsilon_1=D_1\Lambda^2+D_2+\eta_2\Lambda^4/24$  with $C_1,C_2,D_1,D_2,\eta_2$ given in \cite{Luo:2023}, the scaling laws of DWs can be expressed as
\begin{eqnarray}
	D_k^{\alpha\beta}&=&\int_{0}^{\pi}\d k \frac{1}{T(1+\e^{-\kappa/T})(1+\e^{\kappa/T})}\frac{{\widetilde{{q_k^{\alpha}}^{\dr}}}{\widetilde{{q_k^{\beta}}^{\dr}}}(\kappa^{\prime})^2}{\rho^t} =-\frac{e_1\pi^{\frac{1}{2}}}{4(1-2C_1)^{\frac{3}{2}}}T^{\frac{1}{2}}\Li_{\frac{1}{2}}\left(-\e^{\frac{2+C_2}{T}}\right),\label{ck}\\
	D_{\Lambda}^{\alpha\beta}&=&\int_{0}^{\infty}\d \Lambda \frac{1}{T(1+\e^{-\varepsilon_1/T})(1+\e^{\varepsilon_1/T})}\frac{{\widetilde{{q_{\Lambda}^{\alpha}}^{\dr}}}{\widetilde{{q_{\Lambda}^{\beta}}^{\dr}}}(\varepsilon_1^{\prime})^2}{\sigma^t}=-\frac{a\pi^{\frac{1}{2}}}{4D_1^{\frac{3}{2}}}T^{\frac{1}{2}}\Li_{\frac{1}{2}}\left(-\e^{\frac{-D_2}{T}}\right),\label{cs}
\end{eqnarray}
where $a$ is the coefficient in front of $\Lambda^2$ in ${\widetilde{{q_{\Lambda}^{\alpha}}^{\dr}}}{\widetilde{{q_{\Lambda}^{\beta}}^{\dr}}}(\varepsilon_1^{\prime})^2/{\sigma^t}$ in the ground state, and $e_1$ the coefficient in front of $(k-\pi)^2$ in ${\widetilde{{q_{k}^{\alpha}}^{\dr}}}{\widetilde{{q_{k}^{\beta}}^{\dr}}}(\kappa^{\prime})^2/{\rho^t}$ in the ground state.

\textbf{II-IV} ---
For this transition, spin degrees of freedom serves as criticality, thus we can expand $a_n$ around small quantity $\Lambda$
\begin{eqnarray}
	a_1(\sin k-\Lambda)&\approx&a_1(\sin k)-\left[\frac{\pi a_1^2(\sin k)}{u}-\frac{4\pi^2\sin^2 k a_1^3(\sin k)}{u^2}\right]\Lambda^2=-\frac{A_1(k)}{2}-\frac{A_2(k)}{2}\Lambda^2,\\
	a_2(\Lambda-\Lambda^{\prime})&\approx&a_2(\Lambda)-\left[\frac{\pi}{2u}a_2^2(\Lambda)-\frac{\pi^2}{u^2}\Lambda^2a_2^3(\Lambda)\right]{\Lambda^{\prime}}^2=\frac{B_1(\Lambda)}{2}
	+\frac{B_2(\Lambda)}{2}{\Lambda^{\prime}}^2.
\end{eqnarray}
Applying this expansion into eqs.(\ref{dt})(\ref{dct}), we have, for low temperature,
\begin{eqnarray}\left\{\begin{aligned}
		\rho^t&=\frac{1}{2\pi}-\cos k (A_1(k)R_1+A_2(k)R_2),\\
		\sigma^t&=\Lambda^2\left(\frac{R_1}{4\pi u^3}-\frac{3R_2}{8\pi u^5}-S_2\right)+\left(-\frac{R_1}{\pi u}+\frac{R_2}{4\pi u^3}-S_1\right),\label{dd1}
	\end{aligned}\right.\\
	\left\{\begin{aligned}
		\widetilde{q_k^{\dr}}&=\frac{1}{2\pi}-A_1(k)V_1-A_2(k)V_2,\\
		\widetilde{q_{\Lambda}^{\dr}}&=\Lambda^2\left(\frac{V_1}{4\pi u^3}-\frac{3V_2}{8\pi u^5}-W_2\right)+\left(\widetilde{g}-\frac{V_1}{\pi u}+\frac{V_2}{4\pi u^3}-W_1\right),\label{dd2}	
	\end{aligned}\right.
\end{eqnarray}
where $R_{1(2)},S_{1(2)},V_{1(2)},W_{1(2)}$ are defined as
\begin{eqnarray}
	R_1=\int_{0}^{\infty}\d \Lambda \frac{\sigma^t}{1+\e^{\varepsilon_1/T}},R_2=\int_{0}^{\infty}\d \Lambda \frac{\Lambda^2\sigma^t}{1+\e^{\varepsilon_1/T}},S_1=\int_{0}^{\pi}\d k \frac{A_1\rho^t}{1+\e^{\kappa/T}},S_2=\int_{0}^{\pi}\d k \frac{A_2\rho^t}{1+\e^{\kappa/T}},\\
	V_1=\int_{0}^{\infty}\d \Lambda \frac{\widetilde{q_{\Lambda}^{\dr}}}{1+\e^{\varepsilon_1/T}},V_2=\int_{0}^{\infty}\d \Lambda \frac{\Lambda^2\widetilde{q_{\Lambda}^{\dr}}}{1+\e^{\varepsilon_1/T}},W_1=\int_{0}^{\pi}\d k \frac{\cos kA_1\widetilde{q_k^{\dr}}}{1+\e^{\kappa/T}},W_2=\int_{0}^{\pi}\d k \frac{\cos kA_2\widetilde{q_{k}^{\dr}}}{1+\e^{\kappa/T}}.
\end{eqnarray}
Since we intend to relate the critical coefficienst with the ground state quantities, we give the solutions of eqs.(\ref{dd1})(\ref{dd2}) at zero temperature
\begin{eqnarray}\left\{\begin{aligned}
		\rho^0&=\frac{1}{2\pi},\\
		\sigma(0)&=-\frac{1}{2\pi}\int_{0}^{Q}\d kA_1(k)\label{24-d0},
	\end{aligned}\right.\\
	\left\{\begin{aligned}
		{\widetilde{q_k^{\dr}}}^0&=\frac{1}{2\pi},\\
		\widetilde{q_{\Lambda}^{\dr}}(0)&=\widetilde{g}-\frac{1}{2\pi}\int_{0}^{Q}\d k\cos kA_1(k)	\label{24-dc0}.
	\end{aligned}\right.
\end{eqnarray}
In terms of eq.(\ref{cs}) and note that $\varepsilon_1^{\prime}=2D_1\Lambda+\eta_2\Lambda^3/6$, $\widetilde{q_{\Lambda}^{\dr}},\sigma^t$ are given by
\begin{eqnarray}
	\sigma^t&=&\sigma(0)-\frac{\Lambda^2}{2\pi}\int_{0}^{Q}\d kA_2(k),\\
	\widetilde{q_{\Lambda}^{\dr}}&=&\widetilde{q_{\Lambda}^{\dr}}(0)-\frac{\Lambda^2}{2\pi}\int_{0}^{Q}\d k\cos kA_2(k).
\end{eqnarray}
Thus, 
\begin{equation}
	a=\frac{\widetilde{{q_{\Lambda}^{\alpha}}^{\dr}}(0)\widetilde{{q_{\Lambda}^{\beta}}^{\dr}}(0)[\varepsilon_1^{\prime \prime}(0)]^2}{\sigma(0)}.
\end{equation}
Therefore, eq.(\ref{24s}) is proven. 

Turn to the charge sector, the charge part still remains in the low temperature regime, so eq.(\ref{TLL}) is also valid for II-IV transition. Thus we should determine the values of $\rho^t,\widetilde{q_k^{\dr}}$ at finite temperature. In terms of eqs.(\ref{dt}) and (\ref{dct}), we can  obtain
\begin{eqnarray}
	\rho^t&=&\frac{1}{2\pi}-\cos kA_1(k)\int_{0}^{\infty}\d \Lambda \frac{\sigma^t}{1+\e^{\varepsilon_1/T}}\no\\
	&=&\frac{1}{2\pi}+\cos kA_1(k)\int_{0}^{\infty}\d \Lambda  \frac{S_1}{1+\e^{\varepsilon_1/T}}\no\\
	&=&\frac{1}{2\pi}-\cos kA_1(k)\sigma(0)\int_{0}^{\infty}\d \Lambda \frac{1}{1+\e^{\varepsilon_1/T}}\no\\
	&=&\frac{1}{2\pi}-\cos kA_1(k)\sigma(0)f_{\frac{1}{2}},\\
	\widetilde{q_k^{\dr}}&=&\frac{1}{2\pi}-A_1(k)\int_{0}^{\infty}\d \Lambda \frac{\widetilde{q_{\Lambda}^{\dr}}}{1+\e^{\varepsilon_1/T}}\no\\&=&\frac{1}{2\pi}-A_1(k)\int_{0}^{\infty}\d \Lambda \frac{\widetilde{g}-W_1}{1+\e^{\varepsilon_1/T}}\no\\&=&\frac{1}{2\pi}-A_1(k)\widetilde{q_{\Lambda}^{\dr}}(0)f_{\frac{1}{2}},
\end{eqnarray}
where $S_1=-\sigma(0),\widetilde{g}-W_1=\widetilde{q_{\Lambda}^{\dr}}(0),f_{1/2}=\int_{0}^{\infty}\d \Lambda /(1+\e^{\varepsilon_1/T})$. Note that $\kappa^{\prime}=2\sin k$ and substituting above two equations into eqs.(\ref{TLL}) and (\ref{TLLcc1}), we have
\begin{equation}
	D_k^{\alpha\beta}=D^0+2\pi D^0[\cos QA_1(Q)\sigma(0)-A_1(Q)\left(\widetilde{{q_{\Lambda}^{\alpha}}^{\dr}}(0)+\widetilde{{q_{\Lambda}^{\beta}}^{\dr}}(0)\right)]f_{\frac{1}{2}}+\frac{\pi^2T^2}{6}\frac{\partial^2}{\partial \kappa^2}D^0\Big|_{\kappa(k)=0},
\end{equation}
where $D^0=\sin Q/{\pi}$. It can be observed from eqs.(\ref{24-d0}) (\ref{24-dc0}) that $\sigma(0),\widetilde{q_{\Lambda}^{\dr}}(0)$ are functions of charge Fermi points $Q$. Thus $A_1(Q),\cos Q A_1(Q)$ can be related to the derivative of $\sigma(0),\widetilde{q_{\Lambda}^{\dr}}(0)$ w.r.t. $Q$
\begin{equation}
	\sigma^{\prime}(0)=-\frac{1}{2\pi}A_1(Q),{\widetilde{q_{\Lambda}^{\dr}}}^{\prime}(0)=-\frac{1}{2\pi}\cos Q A_1(Q).
\end{equation}
Hence eq.(\ref{24c}) is obtained. In Fig.~\ref{dss}, we show the scaling function for $D^{ss}$ as a function of the magnetic field near the phase transition. The analytical results are in excellent agreement with numerical solutions of the TBA equations.

\begin{figure}[!htb]    
	\centering  
	\includegraphics[scale=0.6]{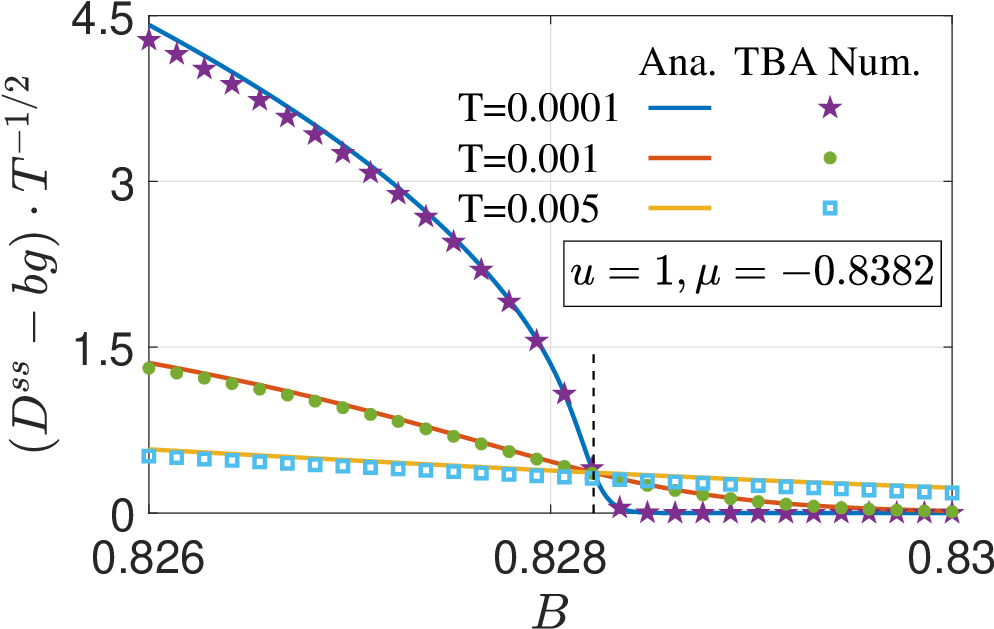} 
	\caption{The scaling behavior of $D^{ss}$ versus magnetic field $B$ for the phase transitoin II-IV with $u=1$, $\mu=-0.8382$. The critical magnetic field for this transition is $B_c=0.8282$, as indicated by the dashed vertical line. The analytical results (solid lines) based on Eqs.~(\ref{24s}) and (\ref{24c}) are in excellent agreement with numerical solutions (symbols) of the TBA equations.}
	\label{dss}  
\end{figure}

\textbf{IV-V} ---
For this transition with charge degrees of freedom vanishing, we can expand $a_1$ around small quantity $\sin k$
\begin{equation}
	a_1(\sin k-\Lambda)\approx a_1(\Lambda)-\left[\frac{\pi a_1^2(\Lambda)}{u}-\frac{4\pi^2\Lambda^2  a_1^3(\Lambda)}{u^2}\right]\sin^2k=\frac{B_3(\Lambda)}{2}+\frac{B_4(\Lambda)}{2}\sin^2k.
\end{equation}
Applying this expansion into eqs.(\ref{dt})(\ref{dct}), we get the equations of density and dressed charge at finite low temperature
\begin{eqnarray}\left\{\begin{aligned}
		\rho^t&=\frac{1}{2\pi}+\cos k (R_3+\sin^2kR_4),\\
		\sigma^t&=S_3B_3(\Lambda)+S_4B_4(\Lambda)-\int_{-\infty}^{\infty}\d \Lambda^{\prime} a_2(\Lambda-\Lambda^{\prime}) \frac{\sigma^t}{1+\e^{\varepsilon_1/T}}\label{45-dT},
	\end{aligned}\right.\\
	\left\{\begin{aligned}
		\widetilde{q_k^{\dr}}&=\frac{1}{2\pi}+V_3+\sin^2kV_4,\\
		\widetilde{q_{\Lambda}^{\dr}}&=\widetilde{g}+W_3B_3(\Lambda)+W_4B_4(\Lambda)-\int_{-\infty}^{\infty}\d \Lambda^{\prime} a_2(\Lambda-\Lambda^{\prime}) \frac{\widetilde{q_{\Lambda}^{\dr}}}{1+\e^{\varepsilon_1/T}}\label{45-dcT},\end{aligned}\right.
\end{eqnarray}
where $R_{3(4)},S_{3(4)},W_{3(4)}$ are defined as
\begin{eqnarray}
	R_3=\int_{0}^{\infty}\d \Lambda \frac{B_3\sigma^t}{1+\e^{\varepsilon_1/T}},R_4=\int_{0}^{\infty}\d \Lambda \frac{B_4\sigma^t}{1+\e^{\varepsilon_1/T}},S_3=\int_{0}^{\pi}\d k \frac{\rho^t}{1+\e^{\kappa/T}},\\S_4=\int_{0}^{\pi}\d k \frac{\sin^2k\rho^t}{1+\e^{\kappa/T}},W_3=\int_{0}^{\pi}\d k \frac{\cos k\widetilde{q_k^{\dr}}}{1+\e^{\kappa/T}},W_4=\int_{0}^{\pi}\d k \frac{\sin^2k\cos k\widetilde{q_{k}^{\dr}}}{1+\e^{\kappa/T}}.
\end{eqnarray}
The solutions at zero temperature are 
\begin{eqnarray}\left\{\begin{aligned}
		\rho(\pi)&=\frac{1}{2\pi}-\int_{-A}^{A}\d \Lambda a_1(\Lambda)\sigma_1(\Lambda),\\
		\sigma(\Lambda)&=\frac{1}{2\pi}\int_{-\pi}^{\pi}\d k a_1(\sin k-\Lambda)-\int_{-A}^{A}\d \Lambda^{\prime} a_2(\Lambda-\Lambda^{\prime})\sigma_1(\Lambda^{\prime})\label{45-d0},
	\end{aligned}\right.\\
	\left\{\begin{aligned}
		{\widetilde{q_k^{\dr}}}(\pi)&=\frac{1}{2\pi}+\int_{-A}^{A}\d \Lambda a_1(\Lambda)\widetilde{q_{\Lambda}^{\dr}}(\Lambda),\\
		\widetilde{q_{\Lambda}^{\dr}}(\Lambda)&=\widetilde{g}-\int_{-A}^{A}\d \Lambda^{\prime} a_2(\Lambda-\Lambda^{\prime})\widetilde{q_{\Lambda}^{\dr}}(\Lambda^{\prime})\label{45-dc0}.
	\end{aligned}\right.
\end{eqnarray}
Contrary to the transition II-IV, the charge degrees of freedom saturates and satisfies the scaling law eq.(\ref{ck}). By analyzing the expressions 
eqs.(\ref{dt}) and (\ref{dct}), we have
\begin{eqnarray}
	\rho^t&=&\rho(\pi)+O(\sin^2 k)+O(\cos k\sin^2k),\\
	\widetilde{q_{k}^{\dr}}&=&\widetilde{q_{k}^{\dr}}(\pi)+O(\sin^2 k).
\end{eqnarray}
Thus,
\begin{equation}
	e_1=\frac{\widetilde{{q_{k}^{\alpha}}^{\dr}}(\pi)\widetilde{{q_{k}^{\beta}}^{\dr}}(\pi)[\kappa^{\prime \prime}(\pi)]^2}{\rho(\pi)}.
\end{equation}
Having known $\kappa^{\prime}\approx -2(1-2C_1)(k-\pi)$ and $1-2C_1=-\kappa^{\prime\prime}(\pi)/2$, we can derive the result eq.(\ref{45c}). To obtain the universal behavior of spin sector, we first capture the properties of $\sigma^t,\widetilde{q_{\Lambda}^{\dr}}$ at finite temperature. The eqs.(\ref{dt}) and (\ref{dct}) can be processed as
\begin{eqnarray}
	\sigma^t&=&-\int_{-A}^{A}\d \Lambda^{\prime} a_2(\Lambda-\Lambda^{\prime}) \sigma(\Lambda^{\prime})+\int_{-\pi}^{\pi}\d k a_1(\sin k-\Lambda) \frac{\rho^t}{1+\e^{\kappa/T}}\no\\
	&=&-\int_{-A}^{A}\d \Lambda^{\prime} a_2(\Lambda-\Lambda^{\prime}) \sigma(\Lambda^{\prime})+\int_{-\pi}^{\pi}\d k a_1(\sin k-\Lambda) \frac{\frac{1}{2\pi}+\cos k (R_3+\sin^2kR_4)}{1+\e^{\kappa/T}}\no\\
	&=&-\int_{-A}^{A}\d \Lambda^{\prime} a_2(\Lambda-\Lambda^{\prime}) \sigma(\Lambda^{\prime})+\int_{-\pi}^{\pi}\d k a_1(\sin k-\Lambda)\frac{1}{2\pi}-\int_{-\pi}^{\pi}\d k a_1(\sin k-\Lambda) \frac{\frac{1}{2\pi}+\cos k R_3}{1+\e^{-\kappa/T}}\no\\
	&=&\sigma^0(\Lambda)-\int_{-A}^{A}\d \Lambda^{\prime} a_2(\Lambda-\Lambda^{\prime}) \sigma^T(\Lambda^{\prime})-\int_{-\pi}^{\pi}\d k a_1(\sin k-\Lambda) \frac{\frac{1}{2\pi}+\cos k R_3}{1+\e^{-\kappa/T}}\no\\
	&=&\sigma^0(\Lambda)-\int_{-A}^{A}\d \Lambda^{\prime} a_2(\Lambda-\Lambda^{\prime}) \sigma^T(\Lambda^{\prime})-\int_{0}^{\pi}\d k \cdot 2a_1(\Lambda) \frac{\frac{1}{2\pi}+\cos k(\frac{1}{2\pi}-\rho(\pi)) }{1+\e^{-\kappa/T}}\no\\
	&=&\sigma^0(\Lambda)-\int_{-A}^{A}\d \Lambda^{\prime} a_2(\Lambda-\Lambda^{\prime}) \sigma^T(\Lambda^{\prime})-2a_1(\Lambda)\rho(\pi)\int_{0}^{\pi}\d k   \frac{1 }{1+\e^{-\kappa/T}}\no\\
	&=&\sigma^0(\Lambda)-\int_{-A}^{A}\d \Lambda^{\prime} a_2(\Lambda-\Lambda^{\prime}) \sigma^T(\Lambda^{\prime})-2a_1(\Lambda)\rho(\pi)k_{\frac{1}{2}}\label{45-m1},
\end{eqnarray}
where the second line used the expression of $\rho^t$ in eq.(\ref{45-dT}). $\sigma^0 (\sigma^T)$ denotes the contribution arising from zero (finite) temperature. The fifth line used the equality $R_3=1/(2\pi)-\rho(\pi)$. It can be observed that eq.(\ref{45-m1}) is an iterative equation of density. Inserting the last temperature term about $k_{\frac{1}{2}}$ into the $\sigma^T$ and iterating over and over, we finally get
\begin{equation}
	\sigma^t=\sigma^0(\Lambda)-2\rho(\pi)k_{\frac{1}{2}}\left\{a_1(\Lambda)-\iint \d \Lambda^{\prime}\d \Lambda^{\prime\prime}a_2(\Lambda-\Lambda^{\prime})(1+a_2)^{-1}(\Lambda^{\prime}-\Lambda^{\prime\prime})a_1(\Lambda^{\prime\prime}) \right\}.
\end{equation}
Considering the value at the Fermi point $A$, we have
\begin{eqnarray}
	\sigma^t|_A&=&\sigma^0(A)-2\rho(\pi)k_{\frac{1}{2}}\left\{a_1(A)-\iint \d \Lambda\d \Lambda^{\prime}a_1(\Lambda)a_2(A-\Lambda^{\prime})(1+a_2)^{-1}(\Lambda-\Lambda^{\prime}) \right\}\no\\
	&=&\sigma^0(A)-2\rho(\pi)k_{\frac{1}{2}}\cdot\iint,
\end{eqnarray}
where we define the term in the final brace as $\iint$. Next, using same way to analyze $\widetilde{q_{\Lambda}^{\dr}}$, we have
\begin{eqnarray}
\widetilde{q_{\Lambda}^{\dr}}&=&	\widetilde{g}-\int_{-A}^{A}\d \Lambda^{\prime} a_2(\Lambda-\Lambda^{\prime}) \widetilde{q_{\Lambda}^{\dr}}+\int_{-\pi}^{\pi}\d k a_1(\sin k-\Lambda)\cos k \frac{\widetilde{q_k^{\dr}}}{1+\e^{\kappa/T}}\no\\
&=&	\widetilde{g}-\int_{-A}^{A}\d \Lambda^{\prime} a_2(\Lambda-\Lambda^{\prime}) \widetilde{q_{\Lambda}^{\dr}}+(\frac{1}{2\pi}+V_3)\int_{-\pi}^{\pi}\d k a_1(\sin k-\Lambda)\cos k \no\\&&-\int_{0}^{\pi}\d k \cdot 2a_1(\Lambda)\cos k \frac{(\frac{1}{2\pi}+V_3)}{1+\e^{-\kappa/T}}\no\\
&=&{\widetilde{q_{\Lambda}^{\dr}}}^0(\Lambda)-\int_{-A}^{A}\d \Lambda^{\prime} a_2(\Lambda-\Lambda^{\prime}) {\widetilde{q_{\Lambda}^{\dr}}}^T(\Lambda^{\prime})-\int_{0}^{\pi}\d k \cdot 2a_1(\Lambda)\cos k \frac{\widetilde{q_k^{\dr}}(\pi)}{1+\e^{-\kappa/T}}\no\\
&=&{\widetilde{q_{\Lambda}^{\dr}}}^0(\Lambda)-\int_{-A}^{A}\d \Lambda^{\prime} a_2(\Lambda-\Lambda^{\prime}) {\widetilde{q_{\Lambda}^{\dr}}}^T(\Lambda^{\prime})+2a_1(\Lambda)\widetilde{q_k^{\dr}}(\pi)k_{\frac{1}{2}}\no\\
&=&{\widetilde{q_{\Lambda}^{\dr}}}^0(\Lambda)+2\widetilde{q_k^{\dr}}(\pi)k_{\frac{1}{2}}\left\{a_1(\Lambda)-\iint \d \Lambda^{\prime}\d \Lambda^{\prime\prime}a_2(\Lambda-\Lambda^{\prime})(1+a_2)^{-1}(\Lambda^{\prime}-\Lambda^{\prime\prime})a_1(\Lambda^{\prime\prime}) \right\},
\end{eqnarray}
where the second line used the expression of $\widetilde{q_k^{\dr}}$ in eq.(\ref{45-dcT}). ${\widetilde{q_{\Lambda}^{\dr}}}^0 ({\widetilde{q_{\Lambda}^{\dr}}}^T)$ denotes the contribution arising from zero (finite) temperature. Thus, the value at the Fremi point is given by
\begin{equation}
	\widetilde{q_{\Lambda}^{\dr}}|_A={\widetilde{q_{\Lambda}^{\dr}}}^0(A)+2\widetilde{q_k^{\dr}}(\pi)k_{\frac{1}{2}}\cdot\iint.
\end{equation}
It can be observed from eqs.(\ref{45-d0})(\ref{45-dc0}) that $\rho(\pi),\widetilde{q_k^{\dr}}(\pi)$ are functions of spin Fermi point $A$. Taking the derivative of $\rho(\pi)$ w.r.t. $A$, we have
\begin{equation}
	\rho^{\prime}(\pi)=-2a_1(A)\sigma^0(A)-\int_{-A}^{A}\d \Lambda a_1(\Lambda)\frac{\d \sigma^0(\Lambda)}{\d A},
\end{equation} 
in which $\frac{\d \sigma^0(\Lambda)}{\d A}$ is given by
\begin{eqnarray}
	\frac{\d \sigma^0(\Lambda)}{\d A}&=&-\sigma^0(A)[a_2(\Lambda-A)+a_2(\Lambda+A)]-\int_{-A}^{A}\d \Lambda^{\prime} a_2(\Lambda-\Lambda^{\prime})\frac{\d \sigma^0(\Lambda^{\prime})}{\d A}\no\\
	&=&-\int_{-A}^{A}\d \Lambda^{\prime}\sigma^0(A)(1+a_2)^{-1}(\Lambda-\Lambda^{\prime})[a_2(\Lambda^{\prime}-A)+a_2(\Lambda^{\prime}+A)].
\end{eqnarray}
Thus, $\rho^{\prime}(\pi)$ can be expressed as
\begin{equation}
	\rho^{\prime}(\pi)=-2\sigma^0(A)\cdot\iint.
\end{equation}
Then,
${\widetilde{q_{k}^{\dr}}}^{\prime}(\pi)$ is given by
\begin{equation}
	{\widetilde{q_{k}^{\dr}}}^{\prime}(\pi)=2{\widetilde{q_{\Lambda}^{\dr}}}^0(A)\cdot\iint.
\end{equation}
Based on eqs.(\ref{TLL}) and (\ref{TLLcc1}), the contribution of the first item is
\begin{eqnarray}
	&&\frac{1}{2\pi}{{q_{\Lambda}^{\alpha}}^{\dr}}(A){{q_{\Lambda}^{\beta}}^{\dr}}(A)v_{\Lambda}(A)\no\\
	&=&\left[{\widetilde{{q_{\Lambda}^{\alpha}}^{\dr}}}^0{\widetilde{{q_{\Lambda}^{\beta}}^{\dr}}}^0+2\left({\widetilde{{q_k^{\alpha}}^{\dr}}}(\pi){\widetilde{{q_{\Lambda}^{\beta}}^{\dr}}}^0(A)+{\widetilde{{q_k^{\beta}}^{\dr}}}(\pi){\widetilde{{q_{\Lambda}^{\alpha}}^{\dr}}}^0(A)\right)\cdot \iint\cdot k_{\frac{1}{2}}\right]\frac{{\varepsilon_1^0}^{\prime}}{\sigma^0-2\rho(\pi)k_{\frac{1}{2}}\cdot\iint}\no\\
	&=&D^0\left[1+k_{\frac{1}{2}}\sum_{e=\alpha, \beta}\frac{\frac{1}{2}\rho(\pi){{q_{k}^{e}}^{\dr}}(\pi)^{\prime}-\rho(\pi)^{\prime}{q_{k}^{e}}^{\dr}(\pi)}{\sigma^0{{q_{\Lambda}^{e}}^{\dr}}^0}\right],
\end{eqnarray}
where ${\varepsilon_1^0}^{\prime}$ is the derivative of $\varepsilon_1^0$ at ground state w.r.t. $\Lambda$, and \begin{equation}
	D^0=\frac{{\varepsilon_1^0}^{\prime}}{\sigma^0}{\widetilde{{q_{\Lambda}^{\alpha}}^{\dr}}}^0{\widetilde{{q_{\Lambda}^{\beta}}^{\dr}}}^0.
\end{equation}
Therefore, the final result eq.(\ref{45s}) is obtained.

\subsection{General relations between DWs and Luttinger parameters}
Here we connect the ground state (i.e., zero temperature) DWs with Luttinger parameters. We will discuss phases IV, II and V. Phase I is vacuum and Phase III represents a full polarized Mott insulator at half filling. These two phases are hence rather trivial.

\textbf{Phase IV:}

In terms of bosonization results, at vanishing magnetic field DWs $D^{cc},D^{ss}$ and susceptibilities $\chi^c,\chi^s$ are related to the Luttinger parameters $K_c,K_s$ via
\begin{eqnarray}
	D^{cc}&=&\frac{K_cv_c}{\pi},\;\chi^c=\frac{2K_c}{\pi v_c}\label{dc-b},\\
	D^{ss}&=&\frac{K_sv_s}{\pi},\;\chi^s=\frac{K_s}{2\pi v_s}\label{ds-b}.
\end{eqnarray}
At finite magnetic field, according to eqs.(\ref{TLL}), we can tell that $D^{cc},D^{ss}$ are combinations of two parts
\begin{eqnarray}
	D^{cc}&=&\frac{1}{2\pi}{q_k^{c,\dr}}^2v_k|_Q+\frac{1}{2\pi}{q_{\Lambda}^{c,\dr}}^2v_{\Lambda}|_A\label{dc-sum},\\
	D^{ss}&=&\frac{1}{2\pi}{q_k^{s,\dr}}^2v_k|_Q+\frac{1}{2\pi}{q_{\Lambda}^{s,\dr}}^2v_{\Lambda}|_A\label{ds-sum}.
\end{eqnarray}
On the other hand, the exact relations about susceptibilities can be derived through the dressed charge matrix
\begin{eqnarray}
	\chi^c|_B&=&\frac{\mathrm{Z}^2_{cc}}{\pi v_c}+\frac{\mathrm{Z}^2_{cs}}{\pi v_s},\label{chi-g-1}\\
	\chi^s|_{\mu}&=&\frac{(\mathrm{Z}_{cc}-2\mathrm{Z}_{sc})^2}{4\pi v_c}+\frac{(\mathrm{Z}_{cs}-2\mathrm{Z}_{ss})^2}{4\pi v_s}. \label{chi-g-2}
\end{eqnarray}
Comparing eqs.(\ref{dc-sum})-(\ref{chi-g-2}) with eqs.(\ref{dc-b})(\ref{ds-b}), it can be observed that, in the presence of magnetic field, the bosonization results eqs.(\ref{dc-b})(\ref{ds-b}) should include crossed contributions from the other degree. By defining cross Luttinger parameters $K_{cs},\,K_{sc}$, DWs can be rewritten as
\begin{eqnarray}
	D^{cc}&=&\frac{K_cv_c}{\pi}+\frac{K_{cs}v_s}{\pi},\;\chi^c=\frac{2K_c}{\pi v_c}+\frac{2K_{cs}}{\pi v_s}\label{dc-bm},\\
	D^{ss}&=&\frac{K_sv_s}{\pi}+\frac{K_{sc}v_c}{\pi},\;\chi^s=\frac{K_s}{2\pi v_s}+\frac{K_{sc}}{2\pi v_c}\label{ds-bm}.
\end{eqnarray}
Thus we can determine the definite expressions of Luttinger parameters (in the following, we omit $|_{Q,A}$ and $q_{k,\Lambda}^{c,s,\dr}$ mean the values at relevant Fermi points)
\begin{eqnarray}
	K_c&=&\frac{{q_k^{c,\dr}}^2}{2}=\frac{\mathrm{Z}^2_{cc}}{2},\;K_s=\frac{{q_{\Lambda}^{s,\dr}}^2}{2}=\frac{(\mathrm{Z}_{cs}-2\mathrm{Z}_{ss})^2}{2},\\
	K_{cs}&=&\frac{{q_{\Lambda}^{c,\dr}}^2}{2}=\frac{\mathrm{Z}^2_{cs}}{2},\;K_{sc}=\frac{{q_{k}^{s,\dr}}^2}{2}=\frac{(\mathrm{Z}_{cc}-2\mathrm{Z}_{sc})^2}{2}.
\end{eqnarray}
Based on above relations and eq.(\ref{TLLcc1}), the cross DWs can be written as 
\begin{equation}
D^{cs}=D^{sc}=\frac{\sqrt{K_cK_{sc}}v_c}{\pi}+\frac{\sqrt{K_sK_{cs}}v_s}{\pi}.
\end{equation}
In the limit of zero magnetic field, for charge transport, the driving term in eq.(\ref{qs-dr}) disappears as $O(1/A^2)$, thus $q_{\Lambda}^{c,\dr}(=\mathrm{Z}_{cs})=0$. In addition, it can be seen from eq.(\ref{q-k}) that $q_{k}^{s,\dr}=0$. Hence, $K_{cs},K_{sc}$ are absent in a system with spin rotation symmetry, leading to the standard bosonization results eqs.(\ref{dc-b})(\ref{ds-b}). In terms of the dressed charge matrix, at vanishing field, the limiting bahavior of $\mathrm{Z}_{ss}$ can be given using the Wiener-Hopf technique, that is $\mathrm{Z}_{ss}=1/\sqrt{2}$. This analysis leads to the known result $K_s=1$, consistent with bosonization result.

\textbf{Phase II}

In the phase II, only charge degree survives. The bosonization of one component Bose gas gives the results
\begin{equation}
	D^{cc}=\frac{K_cv_c}{2\pi},\;\chi^c=\frac{K_c}{\pi v_c}.
\end{equation}
In terms of DW and thermodynamics, $D^{cc},\,\chi^c$ can be expressed as
\begin{equation}
	D^{cc}=\frac{1}{2\pi}{q_k^{c,\dr}}^2v_k|_Q,\;\chi^c=\frac{\mathrm{Z}^2_{cc}}{\pi v_c}.
\end{equation}
Using eq.(\ref{qc-dr}), we can tell $q_k^{c,\dr}=1$. Thus, we have
\begin{equation}
	K_c={q_k^{c,\dr}}^2=\mathrm{Z}^2_{cc}=1,
\end{equation}
corresponding to the result of noninteracting lattice system.

\textbf{Phase V}

In the phase V, charge degree saturates and spin degree plays the role of transport. The bosonization of spin-1/2 Heisenberg model gives the results
\begin{equation}
	D^{ss}=\frac{2K_sv_s}{\pi},\;\chi^s=\frac{K_s}{\pi v_s}.
\end{equation}
The analysis about DW and susceptibility using TBA equations gives the results
\begin{equation}
	D^{ss}=\frac{1}{2\pi}{q_{\Lambda}^{s,\dr}}^2v_{\Lambda}|_A,\;\chi^s=\frac{\mathrm{Z}^2_{ss}}{\pi v_s}.
\end{equation}
Comparing above two equations, we can identify the relation
\begin{equation}
	K_s=\mathrm{Z}^2_{ss}=\frac{{q_{\Lambda}^{s,\dr}}^2}{4}.
\end{equation}
At vanishing magnetic field, Wiener-Hopf technique gives $\mathrm{Z}_{ss}=1/\sqrt{2}$, leading to $K_s=1/2$ which is consistent with isotropic spin chain.


\section{Spin and charge Seebeck effect}
The transport coefficients are defined as
\begin{equation}
\left(\begin{array}{cc}J^c \\ J^s \\ J^{e}\end{array}\right)=\left (\begin{array}{cccc}
\sigma^{cc} &\sigma^{cs}   & \sigma^{ce} \\
\sigma^{sc} &\sigma^{ss} & \sigma^{se} \\
\sigma^{ec} & \sigma^{es} &\sigma^{ee} \\
\end{array}\right)\left(\begin{array}{cc}\nabla \mu \\ \nabla 2 B \\ -\nabla T/T \end{array}\right).\label{sigma}
\end{equation}
The cross conductivity $\sigma^{se} (\sigma^{ce})$ describe the generation of spin (charge) current caused by temperature gradient. In terms of generalized hydrodynamics, the cross Drude weight $D^{se}(\sigma^{ce})$ can be written as 
\begin{equation}
D^{\alpha e}\!=\frac{1}{2T}\sum_{\gamma}\!\!\int \! d\theta_{\gamma} \,\rho_{\gamma}(1-n_{\gamma})\,v_{\gamma}^2  \,{q_{\gamma}^{\alpha}}^{\dr} {q_{\gamma}^{e}}^{\dr},\label{D-dr}
\end{equation} 
where $a=c/s$ denotes dressed charge or spin, ${q_{\gamma}^{e}}^{\dr}$ satisfy
\begin{eqnarray}
{q_{k}^{e}}^{\dr}&=&-2\cos k-\mu-2u-B+\int_{-\infty}^{\infty}\d \Lambda a_1(\sin k-\Lambda) \frac{{q_{\Lambda}^{e}}^{\dr}}{1+\e^{\varepsilon_1/T}},\label{dct1}\\
{q_{\Lambda}^{e}}^{\dr}&=&2B-\int_{-\infty}^{\infty}\d \Lambda^{\prime} a_2(\Lambda-\Lambda^{\prime}) \frac{{q_{\Lambda}^{e}}^{\dr}}{1+\e^{\varepsilon_1/T}}+\int_{-\pi}^{\pi}\d k \cos k a_1(\sin k-\Lambda) \frac{{q_{k}^{e}}^{\dr}}{1+\e^{\kappa/T}}.\label{dct2}
\end{eqnarray}
At ground state, eqs.(\ref{dct1})(\ref{dct2}) are exactly the zero temperature TBA equations, leading to ${q_{k}^{e}}^{\dr}(Q)=0,{q_{\Lambda}^{e}}^{\dr}(A)=0$ with $Q,A$ the Fermi points of ground state.

At low temperature, the universal expression of $D^{\alpha e}$ is given by 
\begin{align}
D^{\alpha e}=\!\!\!\sum_{\gamma=k,\Lambda}\! \frac{1}{2\pi}\!\!\left[{{q_{\gamma}^{\alpha}}^{\dr}{q_{\gamma}^{e}}^{\dr}}v_{\gamma} \!+\!\frac{\pi^2 T^2}{6}\frac{\partial^2({{q_{\gamma}^{\alpha}}^{\dr}{q_{\gamma}^{e}}^{\dr}}v_{\gamma})}{\partial (e_{\gamma}^{\ph})^2}\Big|_{e_{\gamma}^{\ph}=0} \right] .\label{mainTLL}
\end{align}
At ground state, due to ${q_{k}^{e}}^{\dr}(Q)=0,{q_{\Lambda}^{e}}^{\dr}(A)=0$, we have $D^{se}=D^{ce}=0$, indicating a vanishing spin and charge Seebeck effect. In Fig.~\ref{dce}, we plot $D^{ce}$ and $D^{se}$ as functions of chemical potential $\mu$ at different temperatures $T$. Here we separate the contribution from the charge and the spin degrees of freedom. For $D^{ce}$, the two contributions have the same sign; whereas for $D^{se}$, the two contributions have opposite signs in large portions of phase IV.

\begin{figure}[!htb]    
	\centering  
	\includegraphics[scale=0.6]{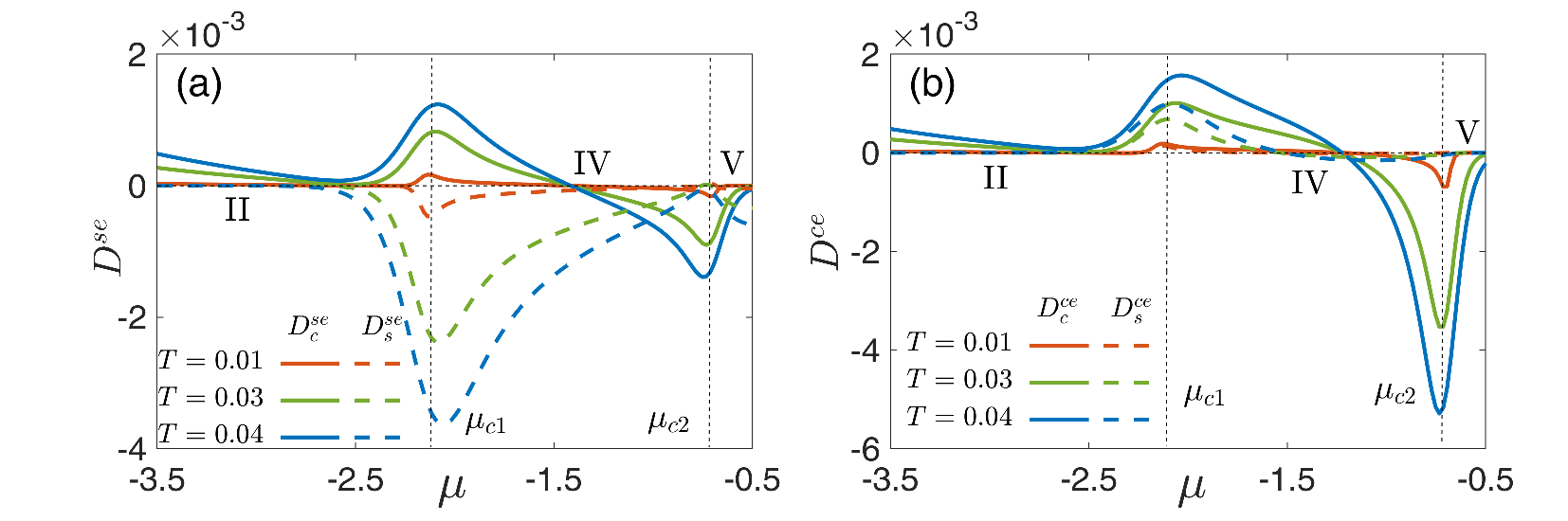} 
	\caption{$D^{se}$ (a) and $D^{ce}$ (b) versus chemical potential $\mu$ for different temperatures $T$. Here the magnetic field and interaction strength are $B=0.5,\,u=1$. The solid (dashed) lines correspond to the contribution from the charge (spin) degrees of freedom. The sum of the two gives the total DW plotted in Fig.~4 of the main text.}
	\label{dce}  
\end{figure}		

An intriguing feature revealed in Fig.~4 of the main text and Fig.~\ref{dce} here is that $D^{se}$ and $D^{ce}$ present pronounced signals close to the phase boundaries. Unlike the scaling behavior of $U(1)$ charges ($D^{cc},D^{ss},D^{sc}$) that follows the scaling of density and magnetization, the spin and charge-thermal transport manifest entropy-like scaling. Considering the spin component $D^{se}_{\Lambda}$, when the spin degree of freedom approaches criticality (e.g., at the II-IV phase transition), its contribution to the Drude weight $D^{se}_{\Lambda}$ can be calculated as
\begin{eqnarray}
D_{\Lambda}^{se}&=&\int_{0}^{\infty}\d \Lambda \frac{1}{T(1+\e^{-\varepsilon_1/T})(1+\e^{\varepsilon_1/T})^2}\frac{{{q_{\Lambda}^{s}}^{\dr}}{{q_{\Lambda}^{e}}^{\dr}}(\varepsilon_1^{\prime})^2}{4\pi^2\sigma}\no\\
&=&\int_{0}^{\infty}\d \Lambda \frac{(a_1\Lambda^2+a_2)}{T(1+\e^{-\varepsilon_1/T})(1+\e^{\varepsilon_1/T})}\frac{{{q_{\Lambda}^{s}}^{\dr}}{{q_{\Lambda}^{e}}^{\dr}}(\varepsilon_1^{\prime})^2}{4\pi^2}\no\\
&=&\int_{0}^{\infty}\d \Lambda \frac{(b_1\Lambda^4+b_2\Lambda^2)\varepsilon_1}{T(1+\e^{-\varepsilon_1/T})(1+\e^{\varepsilon_1/T})}\no\\
&=&\int_{0}^{\infty}\d \Lambda \frac{(b_1\Lambda^4+b_2\Lambda^2)(D_1\Lambda^2+D_2)}{T(1+\e^{-\varepsilon_1/T})(1+\e^{\varepsilon_1/T})}\no\\
&=&\int_{0}^{\infty}\d \Lambda \frac{(b_2D_2\Lambda^2+b_2D_1\Lambda^4)}{T(1+\e^{-\varepsilon_1/T})(1+\e^{\varepsilon_1/T})}\no\\
&=&-\frac{b_2D_2\pi^{\frac{1}{2}}}{4D_1^{\frac{3}{2}}}T^{\frac{1}{2}}\Li_{\frac{1}{2}}\left(-\e^{\frac{-D_2}{T}}\right)-\frac{3b_2\pi^{\frac{1}{2}}}{8D_1^{\frac{3}{2}}}T^{\frac{3}{2}}\Li_{\frac{3}{2}}\left(-\e^{\frac{-D_2}{T}}\right),\label{cscs}
\end{eqnarray}
where $\varepsilon_1=D_1\Lambda^2+D_2$  with $C_1,C_2,D_1,D_2$ given in \cite{Luo:2023}. In the critical region $\sigma$ is in the form of the quadratic function of rapidity $\Lambda$. In the second line, we used the denotation $\sigma(1+\e^{\varepsilon_1/T})\equiv 1/(a_1\Lambda^2+a_2)$ with $a_1,a_2$ coefficents; in the third line, ${q_{\Lambda}^{e}}^{\dr}\approx \partial \left(\beta \varepsilon_1\right)/\partial \beta \approx \varepsilon_1$
and we donote the remaining coefficients as $(b_1\Lambda^4+b_2\Lambda^2)$; in the fifth line, the term $b_1D_2\Lambda^4$ is neglected as it represents  a small quantity. Rearranging the eq.(\ref{cscs}) as
\begin{equation}
D_{\Lambda}^{se}=-\frac{b_2\pi^{\frac{1}{2}}}{4D_1^{\frac{3}{2}}}\left\{D_2T^{\frac{1}{2}}\Li_{\frac{1}{2}}\left(-\e^{\frac{-D_2}{T}}\right)+\frac{3}{2}T^{\frac{3}{2}}\Li_{\frac{3}{2}}\left(-\e^{\frac{-D_2}{T}}\right)\right\}.\label{se}
\end{equation}
Note that the entropy in the II-IV phase transition satisfies
\begin{equation}
s=s_0+\pi^{\frac{1}{2}}\sigma_1(0)\left(\frac{\varepsilon^{''}_1(0)}{2}\right)^{-\frac{1}{2}}
\left\{D_2T^{-\frac{1}{2}}\Li_{\frac{1}{2}}\left(-\e^{\frac{-D_2}{T}}\right)+\frac{3}{2}T^{\frac{1}{2}}\Li_{\frac{3}{2}}\left(-\e^{\frac{-D_2}{T}}\right)\right\}.\label{s}
\end{equation}
Compare eqs. (\ref{se}) and (\ref{s}), it can be shown that $D_{\Lambda}^{se}$ and $Ts$ display same scaling behavior. Based on eq.(\ref{se}), $D_{\Lambda}^{se}$ can reach extreme point within quantum critical region. From the result of $\partial D_{\Lambda}^{se}/\partial \mu=0$, the condition for
determining the extreme point reads
\begin{equation}
\frac{1}{2}\Li_{\frac{1}{2}}\left(-\e^{\frac{-D_2}{T}}\right)+\frac{D_2}{T}\Li_{-\frac{1}{2}}\left(-\e^{\frac{-D_2}{T}}\right)=0.
\end{equation}
It is found that, to a good approximation, $-D_2/T\approx 1.3117$.
This principle facilitates the design of highly sensitive temperature sensors in quantum critical regions. When a temperature gradient is applied, spin Seebeck effect generates a spin current. Upon entering an adjacent metal layer, this spin flow is converted into a measurable voltage signal via the inverse spin Hall effect, thereby enabling electrical detection of thermal signals.

\section{Numerical simulation}

The DWs are computed by examining the dynamics of a quenched spin-${1}/{2}$ Fermi gas in a 1D optical lattice. Dynamics is governed by the 1D Fermi--Hubbard Hamiltonian 
\begin{equation}
	H^{c,s}(t) =- \sum^L_{j=1, a=\uparrow,\downarrow}\left(c_{j+1,a}^{\dagger} c_{j, a}+{\rm H.c.}\right)+u \sum_{j=1}^{L} (1-2n_{j \uparrow}) (1-2n_{j \downarrow}) + V^{c,s}(t) \,, 
\end{equation}
where $c_{j,a}^\dagger$ ($c_{j,a}$) is the creation (annihilation) operator and $n_j=\sum_{a}n_{j,a}$ is the local density of a fermionic atom with spin $a$ ($a=\uparrow$ or $\downarrow$) at site $j$ on a 1D lattice of length $L$. $u$ is the on-site interaction. 
The density and the spin dependent gradient potentials coupling to the charge and spin degrees of freedom respectively are defined as $V^{c,s}(t) = V^{c,s}\Theta(t)$, where the Heaviside function $\Theta(t) = 1$ for $t>0$ else $0$ and the coefficients
\begin{align}
    V^{c} = \Delta_c \sum_{j=1}^{L}n_j x_j; \quad V^{s} = \Delta_s \sum_{j=1}^{L}(2S^z_j) x_j\,,
\end{align}
where $x_j = {(L+1)}/{2}-j$ denotes the position of site $j$ from the center of the system and $\Delta_c$ $\&$ $\Delta_s$ are the strength of the gradient potentials. The introduction of such a weak constant force induces charge density and spin magnetization currents that are used to extract the DWs. 

\textit{Generalized effective spin-chain (GESC) model}: Analysis is done at strong on-site interaction using the generalized effective spin-chain model \cite{basak_generalized_2023}, which enables higher accuracy by allowing the consideration of larger system sizes. The charge degrees of freedom in the model is governed by the spinless Fermi-Hubbard Hamiltonian with an effective gradient potential that accounts for the effect of the charge- or spin-dependent gradient potential on the charge degrees of freedom
\begin{equation}
	H^{c,s}_{c}(t) =- \sum^L_{j=1}\left(f_{j+1}^{\dagger} f_{j}+{\rm H.c.}\right)+ V^{c,s}_{\rm{eff},c}(t)\,, 
\end{equation}
where $f_{j}^\dagger$ ($f_{j}$) is the creation (annihilation) operator of the spinless fermion. The effective gradient potential
\begin{align}
    V^{c}_{\rm{eff},c}(t) = \Theta(t) \Delta_c \sum_{j=1}^{L}n^f_j x_j; \quad V^{s}_{\rm{eff},c} (t)= 2\Theta(t) \Delta_s \left(\dfrac{m}{n_c}\sum_j x_j n^f_j +\left(\dfrac{m}{n_c}-\dfrac{1}{2}\right)\sum_j |x_j| n^f_j\right)
\end{align}
where $n^f_j = f_{j}^\dagger f_{j}$ is the number operator for spinless fermions, $m$ is the magnetization and $n_c = N/L$ is the site occupancy. The effective spin dependent gradient potential is obtained by computing the collective magnetization dependent effect on the charge degrees of freedom when the spin degrees of freedom couple to the potential. The spin state is governed by the spin-chain Hamiltonian with an effective gradient potential
\begin{align}
    H^{c,s}_{SC}(t) =-\dfrac{1}{4u}\sum_l\mathcal{C}^{c,s}_l(t)\left(I-\mathcal{E}_{l,l+1}\right) + V^{c,s}_{\rm{eff},s}(t)\,,
\end{align}
where  $\mathcal{E}_{l,l+1}$ an exchange operator acting on spins $l$ and $l+1$ and exchanging them and the coupling coefficient ($\mathcal{C}_l$) for spin $l$ is given by
\begin{align}
    \mathcal{C}^{c,s}_l(t) =  \langle \psi ^{c,s} (t) |{\sum_{j=1}^{L-1}\delta_{\sum_{i=1}^{j-1}n^f_i,l-1}\left(2n^f_jn^f_{j+1}-f_{j+2}^{\dagger}n^f_{j+1}f_j^{\vphantom{\dagger}}-f_{j-1}^{\dagger}n^f_jf_{j+1}^{\vphantom{\dagger}}\right)}{\varphi^{c,s}(t)} \,.
\end{align}
$\varphi^{c,s}(t)$ is the charge wavefunction. The effective gradient potential
\begin{align}
    V^{c}_{\rm{eff},s}(t) = 0; \quad V^{s}_{\rm{eff},c}(t) = \Theta(t)\Delta_s \sum_l \mathcal{D}^{c,s}_l(t) \sigma^z_l\,,
\end{align}
where $\sigma^z_l = n_{l,\uparrow}-n_{l,\downarrow}$ and coefficient 
\begin{align}
    \mathcal{D}^{c,s}_l(t) =  \langle \psi ^{c,s} (t) |{\sum_i \delta_{\sum_{j=1}^{i-1}n^f_j,l-1} x_i n^f_i}{\varphi^{c,s}(t)}\,,
\end{align}
For the numerics, we consider open boundary conditions (OBC) and a canonical ensemble with fixed numbers in each spin component, for which reason, we don't need to include the chemical potential and the magnetic field terms in the Hamiltonian. Conserved quantities charge density $\hat{n} = \sum_j \left(n_{j,\uparrow}+n_{j,\downarrow}\right) = \sum_j n_j$ and spin magnetization $2\hat{S}^z = \sum_j \left(n_{j,\uparrow}-n_{j,\downarrow}\right) = \sum_j (2S^z_j)$ are examined to study the charge and spin transport of this system. 

\textit{Charge and spin currents}: The local charge density ($j_c(t)$) and spin magnetization ($j_s(t)$) currents are obtained at the system center by computing the imbalance in the density and the spin magnetization between the right and left of the system center \cite{schuttelkopf_characterising_2024}. 
\begin{align}
    \begin{split}
        & j_c(t) = \dfrac{1}{2}\dfrac{d\Delta N(t)}{dt} \approx \dfrac{\Delta N(t)}{t}
        \\& j_s(t) = \dfrac{1}{2}\dfrac{d\Delta 2S^z(t)}{dt} \approx \dfrac{\Delta 2S^z(t)}{t} 
    \end{split}
    \label{eqn:jcs}
\end{align}
where $\Delta N(t) = \sum^L_{j=L/2+1}n_j(t)-\sum^{L/2}_{j=0}n_j(t)$ is the density imbalance and $\Delta 2S^z(t) = \sum^L_{j=L/2+1}2s_j^z(t)-\sum^{L/2}_{j=0}2s_j^z(t)$ is the spin magnetization imbalance. The second equality in the expressions of $j_c(t)$ and $j_s(t)$ holds true for a linear growth of the current observed at long times. 

\textit{Drude Weights}: For weak constant force, the DWs are obtained by examining the dissipationless current displaying the time-asymptotic growth of the local current \cite{schuttelkopf_characterising_2024}
\begin{align}
     D^{\alpha \beta} = \lim_{\Delta_\alpha \rightarrow 0} \lim_{t \rightarrow \infty} \,\frac{j_\beta(t)}{t \Delta_\alpha},
\end{align}
where $j_b\left(z,t\right)$ represents the local current of operator $b$ at position $z$ and $\Delta_a$ is the gradient of the potential coupled to operator $a$. 

Assuming weak linear gradient potentials ($V^{c,s}$), the DWs from the local charge density and spin magnetization currents at position $x_j =0$
\begin{align}
    \begin{split}
        & D^{cc} = \lim_{t \rightarrow \infty}\dfrac{j_c(t,\Delta_c;\Delta_s=0)}{t\Delta_c}, \quad  D^{cs} = \lim_{t \rightarrow \infty}\dfrac{j_s(t,\Delta_c;\Delta_s=0)}{t\Delta_c},
        \\& D^{sc} = \lim_{t \rightarrow \infty}\dfrac{j_c(t,\Delta_s;\Delta_c=0)}{t\Delta_s}, \quad  D^{ss} = \lim_{t \rightarrow \infty}\dfrac{j_s(t,\Delta_s;\Delta_c=0)}{t\Delta_s}.
    \end{split}
    \label{eqn:dws}
\end{align}

\textit{Numerical Parameters}: The initial state and the dynamics are obtained from the GESC model using time-dependent variational principle method (developed with input from \cite{mendl_pytenet_2018}) for a system size of $L=40$, number of particles $N=20$, potential gradients $\Delta_c=\Delta_s=0.1$, interaction strength $u = 15$ and OBC (box potential $V_w = 100$). Time dependent variational principle (TDVP), like tDMRG, simulates the time evolution of quantum systems using matrix product states. However, it differs by using tangent space for simulation, instead of the Lie-Trotter decomposition used in the local Krylov method. This makes TDVP particularly useful for restricting the time evolution to a specific set of matrix-product states. DWs are computed for different magnetization ($m = {(n_{\uparrow}-n_{\downarrow})}/{(2L)}$; $n_{\sigma} = \sum_j n_{\sigma,j}$). The finite system size sets the maximal time ($t = t_{\rm{max}}=5 < \infty$) of the numerical simulation after which the spin and charge motion opposite to the induced current arise due to the box potential for open boundary conditions. This results in the density and spin magnetization imbalances to deviate from a quadratic growth and the extracted currents from a linear growth beyond the maximal time.

\begin{figure}
\centering
\includegraphics[width=\linewidth]{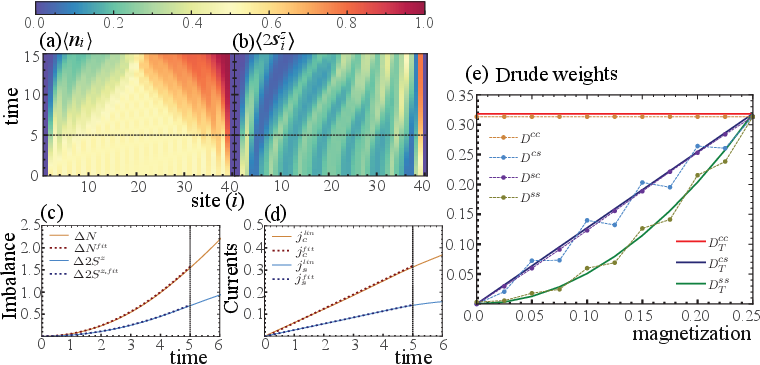}
\caption{\label{fig:schematic}(a-d) Numerical extraction of the charge and spin currents in the presence of density dependent gradient potential ($V^c$) at magnetization $m = 0.1$. Evolution of the local (a) density $(n_i)$ and (b) spin magnetization $(2s^z_i)$ as a function of site $(i)$ and time. (c) Imbalance in density and spin magnetization with quadratic fits. (d) Charge and spin currents obtained from the quadratic fits and linear approximations. (e) Scaled Drude weights obtained numerically as a function of magnetization with the theory predictions ($D^T_{cc}\approx \sin(\pi n_c)/\pi,D^T_{cs} \approx 2m\sin(\pi n_c)/n_c\pi ,D^T_{ss}\approx 4m^2\sin(\pi n_c)/n^2_c\pi$). System parameters: $L=40$, $N=20$, $u=15$, $\Delta_c =0.1$, $\Delta_s =0.1$, GESC model and open boundary conditions. Time $(t)$ is scaled with tunneling. Black dashed line represents
the cutoff time.}	
\end{figure}

\textit{Results}: To illustrate how Drude weights are computed, the evolution of fermions trapped in an optical lattice (OBC) and quenched with a linear density-dependent potential is considered in Figs.~\ref{fig:schematic} that shows large$u$ results based on GESC.  Analyzing the local density ($\langle{n_i}\rangle$) and spin magnetization ($\langle{2S^z_i}\rangle$) shown in Figs.~\ref{fig:schematic}(a) and (b), it is observed that the density and spin magnetization experience the walls at time $t \approx 5$. Beyond this time, motion against the induced current is seen, resulting in oscillations. Thus, for the process of computing the Drude weights the analysis is restricted to time $t=5$. Fig.~\ref{fig:schematic}(c) presents the imbalance of the density ($\Delta N(t)$) and spin magnetization ($2S^z(t)$) about the system center. The imbalances exhibit quadratic growth, verified by numerical quadratic fits. Using quadratic fits, the charge and spin currents ($j^{\rm{fit}}_c,j^{\rm{fit}}_s$) are calculated as the first equality in Eq.~(\ref{eqn:jcs}). The currents obtained demonstrate a linear growth as shown in Fig.~\ref{fig:schematic}(d). Given linear growth, linear approximations of the currents ($j^{\rm{lin}}_c$, $j^{\rm{lin}}_s$ as the second equality in Eq.~(\ref{eqn:jcs})) are sufficient to calculate the DW. This is further supported by the good agreement between the linear approximations of the current and the ones from the quadratic fit. DWs computed using the linear approximations of the currents at time $t = t_{\rm{max}}=5$,  $n_c D^{cc} =  n_cj^{\rm{lin}}_c/5\Delta_c \approx 0.31$ and $n_c D^{cs} = n_cj^{\rm{lin}}_s/5\Delta_c \approx 0.14$. DWs obtained using the GESC model require an overall scaling factor $\approx 1/2$ to align well with the theoretical predictions. Fig.~\ref{fig:schematic}(e) shows the scaled DWs computed for different values of magnetization. Additionally, the deviations of $D^{cs}$ and $D^{ss}$ from their expected linear and quadratic dependence on magnetization vanish when magnetizations corresponding to even numbers of spin-up and spin-down particles are considered. This odd-even symmetry arises because of the finite system size and is expected to disappear in the thermodynamic limit. In general, the computed DWs show excellent qualitative agreement and strong quantitative agreement with the theoretical predictions as shown in Fig.~\ref{fig:schematic}(e).  



\end{widetext}


\begin{thebibliography}{10}

\bibitem{Kubo:1957}
R. Kubo, J. Phys. Soc. Jpn, {\bf 12} 570 (1957).

\bibitem{Kubo:book}
R. Kubo, M. Toda and N. Hashitsume, 
{\it Statistical physics II: nonequilibrium statistical mechanics} (Springer,
Berlin, 1991).

\bibitem{Callaway:1959}
J. Callaway,
Phys. Rev. {\bf 113} 1046 (1959).

\bibitem{Hartnoll:2012}
S. A. Hartnoll and D. M. Hofman,
Phys. Rev. Lett. {\bf 108} 241601 (2012).

\bibitem{Maznev:2014}
A. A. Maznev and O. B. Wright, 
Am. J. Phys. {\bf 82} 1062 (2014).

\bibitem{Rosch:2000}
A. Rosch and N. Andrei, 
Phys. Rev. Lett. {\bf 85} 1092 (2000).

\bibitem{Nardis:2019}
J. De Nardis, D. Bernard and B. Doyon 
SciPost Phys. {\bf 6} 49 (2019).

\bibitem{Mazur:1969}
P. Mazur,
Physica {\bf 43}, 533  (1969).
\bibitem{Zotos:1997}
X. Zotos, F. Naef and P. Prelovsek,
Phys. Rev. B {\bf 55}, 11029  (1997).
\bibitem{Sirker:2011}
J. Sirker, R. G. Pereira and I. Affleck,
Phys. Rev. B {\bf 83}, 035115 (2011).
\bibitem{Sirker:2020}
J. Sirker,
SciPost Phys. Lect. Notes {\bf 17},  (2020).

\bibitem{Scalapino:1992}
D. J. Scalapino, S. R. White, and S. C. Zhang,
Phys. Rev. Lett. {\bf 68}, 2830 (1992).


\bibitem{Kohn:1964}
W. Kohn,
Phys. Rev. {\bf 133} A171 (1964).
\bibitem{Shastry:1990}
B. S. Shastry and B. Sutherland,
Phys. Rev. Lett. {\bf 65} 243 (1990).

\bibitem{Castella:1995}
H. Castella, X. Zotos and P. Prelovsek, Phys. Rev. Lett. {\bf 74} 972 (1995).
\bibitem{Zotos:1996}
X. Zotos and P. Prelovsek,
Phys. Rev. B {\bf 53} 983 (1996).
\bibitem{Narozhny:1998}
B. N. Narozhny, A. J. Millis, and N. Andrei
Phys. Rev. B {\bf 58} R2921(R) (1998).

\bibitem{Zotos:1990}
X. Zotos, 
Phys. Rev. Lett. {\bf 82} 1764 (1999).
\bibitem{Zotos:2017}
X. Zotos, 
J. Stat. Mech. (2017) 103101.

H. Watanabe and M. Oshikawa, 
Phys. Rev. B {\bf 102} 165137  (2020).
M. Fava, S. Biswas, S. Gopalakrishnan, R. Vasseur and S. A. Parameswaran,
Proc. Natl Acad. Sci.
{\bf 118} e2106945118  (2021).
Y. Tanikawa, K. Takasan and H. Katsura, 
Phys. Rev. B {\bf 103} L201120  (2021).
Y. Tanikawa and H. Katsura, 
Phys. Rev. B {\bf 104} 205116  (2021).


\bibitem{Castro:2016}
O. A. Castro-Alvaredo, B. Doyon and T. Yoshimura
Phys. Rev. X {\bf 6} 041065 (2016).
\bibitem{Bertini:2016}
B. Bertini, M. Collura, J. De Nardis and M. Fagotti 
Phys. Rev. Lett. {\bf 117} 207201 (2016).
\bibitem{Ilievski:2017}
E. Ilievski and J. De Nardis 
Phys. Rev. B {\bf 96} 081118(R) (2017).

%
\bibitem{Doyon:2017}
B. Doyon and H. Spohn
SciPost Phys. {\bf 3} 39 (2017).
\bibitem{Fava:2020}
M. Fava, B. Ware, S. Gopalakrishnan, R. Vasseur, and S. A. Parameswaran 
Phys. Rev. B {\bf 102} 115121  (2020).
\bibitem{Bulchandani:2018}
V. B. Bulchandani, R. Vasseur, C. Karrasch, and J. E. Moore 
Phys. Rev. B {\bf 97} 045407  (2018).

\bibitem{Nardis:2018}
J. De Nardis, D. Bernard, and B. Doyon
Phys. Rev. Lett. {\bf 121} 160603 (2018).
\bibitem{Panfil:2019}
M. Panfil and J. Pawełczyk 
SciPost Phys. Core {\bf 1}, 002 (2019)
\bibitem{Doyon:2022}
B. Doyon
J Stat Phys {\bf 186} 25 (2022).
\bibitem{Ilievski:2018}
E. Ilievski, J. De Nardis, M. Medenjak, and T. Prosen
Phys. Rev. Lett. {\bf 121} 230602 (2018).


\bibitem{Gopalakrishnan:2018}
S. Gopalakrishnan, D. A. Huse, V. Khemani, and R. Vasseur,  Phys. Rev. B {\bf 98}, 220303(R) (2018).
\bibitem{Gopalakrishnan:2019}
S. Gopalakrishnan and R. Vasseur,  Phys. Rev. Lett. {\bf 122}, 127202 (2019).

\bibitem{Caux:2019}
J.-S. Caux, B. Doyon, J. Dubail, R. Konik and T. Yoshimura,  SciPost Phys.  {\bf 6}, 70  (2019).
\bibitem{Schemmer:2019}
M. Schemmer, I. Bouchoule, B. Doyon, and J. Dubail,  Phys. Rev. Lett. {\bf 122}, 090601  (2019).
\bibitem{Bastianello:2019}
A. Bastianello, V. Alba, and J.-S. Caux,  Phys. Rev. Lett. {\bf 123}, 130602  (2019).
\bibitem{Malvania:2021}
N. Malvania, Y. Zhang, Y. Le, J. Dubail, M. Rigol and D. S. Weiss,  Science {\bf 373}, 1129 (2021).
\bibitem{Ruggiero:2020}
P. Ruggiero, P. Calabrese, B. Doyon, and J. Dubail,  Phys. Rev. Lett. {\bf 124}, 140603  (2020).

\bibitem{Guan:2022}X.-W. Guan and P. He, Rep. Prog. Phys. {\bf 85}, 114001 (2022).

\bibitem{Rmer:1995}
R. A. R{\"o}mer and A. Punnoose, 
Phys. Rev. B {\bf 52} 14809 (1995).


					

\bibitem{Nozawa:2020}
Y. Nozawa and H. Tsunetsugu, 
Phys. Rev. B {\bf 101} 035121 (2020).

\bibitem{Nozawa:2021}
Y. Nozawa and H. Tsunetsugu, 
Phys. Rev. B {\bf 103} 035130 (2021).
\bibitem{Karrasch:2014}
C. Karrasch, D. M. Kennes and J. E. Moore, 
Phys. Rev. B {\bf 90} 155104 (2014).
\bibitem{Fujimoto:1998}
S. Fujimoto and N. Kawakami, 
J. Phys. A: Math. Gen. {\bf 31} 465 (1998).

\bibitem{Lieb:1968}
E.~H. Lieb and F.~Y. Wu,  Phys. Rev. Lett. {\bf 20}, 1445 (1968).
\bibitem{Lieb:2003}
E.~H. Lieb and F.~Wu, Phys. A {\bf 321}, 1 (2003).

\bibitem{Ess05}F. H. L. Essler, H. Frahm, F. G\"{o}hmann, A. Kl\"{u}mper and V. E. Korepin,
{\it The One-Dimensional Hubbard Model} (Cambridge University Press, Cambridge, 2005).

\bibitem{Nardis:2020}
J. De Nardis, S. Gopalakrishnan, E. Ilievski, and R. Vasseur, Phys. Rev. Lett. {\bf 125}, 070601 (2020).








\bibitem{Tajik:2024} P. Schüttelkopf, M. Tajik, N. Bazhan, F. Cataldini, S.-C. Ji, J. Schmiedmayer and F. Møller, 	arXiv:2406.17569.



\bibitem{Guan:1998}X.-W. Guan and S.-D. Yang, Nucl. Phys. B {\bf 512}, 601 (1998). 

\bibitem{Urichuk:2022}
A. Urichuk, A. Klümper, and J. Sirker, 
Phys. Rev. Lett. {\bf 129} 096602  (2022).


\bibitem{JJL:SM}
Supplemental Material. Here, we present brief derivations of charge, spin and cross DWs for the 1D Hubbard model with an arbitrary magnetic field at low temperatures. In particular, explicit relations among the  Luttinger parameters, dressed charges and DWs are obtained in detail. 

\bibitem{Luo:2023-PRB}J.-J. Luo, H. Pu, X.-W. Guan, Phys. Rev. B {\bf 107}, L201103 (2023). 

\bibitem{Luo:2023}
J.-J. Luo, H. Pu, X.-W. Guan, Rep. Prog. Phys.  {\bf 87}, 117601 (2024).

\bibitem{Boll:2016} M. Boll, T. A.  Hilker, G. Salomon, A. Omran, J. Nespolo, Pollet L, I. Bloch, and C. Gross, {Science} {\bf 353}, 1257 (2016).
\bibitem{Hilker:2017} T.  A. Hilker, G. Salomon,  F. Grusdt, A. Omran, M. Boll, E. Demler, I. Bloch, and C. Gross,  {Science} {\bf 357},  484 (2017). 
\bibitem{Vijayan:2020} J. Vijayan, P. Sompet, G. Salomon, J. Koepsell, S. Hirthe, A. Bohrdt, F. Grusdt, I. Bloch, and C. Gross, {Science} {\bf 367}, 186 (2020). 

\bibitem{Spar:2022} B. M. Spar, E. Guardado-Sanchez, S.  Chi, Z. Z. Yan and W. S. Bakr, Phys. Rev. Lett. {\bf 128}, 223202 (2022).


\bibitem{Giamarchi:1997} T. Giamarchi, Physica B Condens. Matter {\bf 230}, 975 (1997).
\bibitem{Giamarchi-book} T. Giamarchi,  {\em Quantum physics in one dimension} (Oxford: Oxford University Press, 2004). 


\bibitem{Geballe1955} T. H. Geballe and G. W. Hull, Phys. Rev. {\bf 98}, 940 (1955).
\bibitem{Herwaarden1986} A. W. Van Herwaarden and P. M. Sarro, Sensors and Actuators {\bf 10}, 321 (1986).
\bibitem{Goldsmid2010}H. J. Goldsmid, {\em Introduction to Thermoelectricity, Berlin: Springer, 2010.   }

\bibitem{adachi2013theory} H. Adachi, K.-i. Uchida, E. Saitoh and S. Maekawa, Rep. Prog. Phys. {\bf 76}, 036501 (2013).
\bibitem{Uchida2008} K. Uchida, S. Takahashi, K. Harii, J. Ieda, W. Koshibae, K. Ando, S. Maekawa and E. Saitoh, Nature {\bf 455}, 778 (2008).
\bibitem{Jaworski2010} C. M. Jaworski, J. Yang, S. Mack, D. D. Awschalom, J. P. Heremans and R. C. Myers, Nature Mater. {\bf 9}, 898 (2010).
\bibitem{Barfknecht2021} R. E. Barfknecht, A. Foerster, N. T. Zinner and A. G. Volosniev, Communications Physics, 4:252 (2022). 
%

\bibitem{Moore:2007} R. G. Moore, J. Zhang, V. B. Nascimento, R. Jin, J. Guo, G.T. Wang, Z. Fang, D. Mandrus and E. W. Plummer, Science {\bf 318}, 615 (2007).
\bibitem{Koretsune:2007}
T. Koretsune, Y. Motome and A. Furusaki, J. Phys. Soc. Japan {\bf 76}, 074719 (2007).










\bibitem{Bardon:2014}A.  Bardon, S. Beattie, C. Luciuk, W. Cairncross, D. Fine, N. S. Cheng, G. J. A. Edge, E. Taylor, Shizhong Zhang, S. Trotzky, J. H. Thywissen, Science {\bf 344}, 722 (2014).
	
\bibitem{Matthew:2019} M. A. Nichols, L. W. Cheuk, M. Okan, T. R. Hartke, E. Mendez, T. Senthil, E. Khatami, H. Zhang and M. W. Zwierlein, Science, {\bf 363}, 383 (2019).	
\bibitem{Amico:Atomtronic}L. Amico, D.  Anderson, M.  Boshier, J.-P. Brantut, L. C.  Kwek, A.  Minguzzi, and W. von Klitzing	Rev. Mod. Phys. {\bf 94}, 041001 (2022).	
				
\end{thebibliography}

\begin{thebibliography}{199}
	
	\bibitem{Guan:1998}X.-W. Guan and S.-D. Yang, Nucl. Phys. B {\bf 512}, 601 (1998).
	
	
	\bibitem{Luo:2023}
	J.-J. Luo, H. Pu and X.-W. Guan, Rep. Prog. Phys. {\bf 87}, 117601 (2024).

    \bibitem{basak_generalized_2023} S. Basak and H. Pu, Phys. Rev. A {\bf 108}, 063315  (2023).

    \bibitem{schuttelkopf_characterising_2024} P. Sch\"{u}ttelkopf, M. Tajik, N. Bazhan, F. Cataldini, S.-C. Ji, J. Schmiedmayer, F. M{\o}ller, arXiv:2406.17569 (2024).

    \bibitem{mendl_pytenet_2018}C. Mendl, J. Open Source Software {\bf 3}, 948 (2018).

	
	
\end{thebibliography}
\end{document}